\documentclass[usenatbib]{mnras}

\usepackage{newtxtext,newtxmath}

\usepackage[T1]{fontenc}
\usepackage{ae,aecompl}

\usepackage{graphicx}	
\usepackage{amsmath}	
\usepackage{amssymb}	
\usepackage{multicol}
\usepackage{hyperref}

\title[ELAIS N1 with uGMRT]{ Detailed study of the  ELAIS N1 field with the uGMRT - I. Characterizing the 325 MHz foreground for redshifted 21 cm observations}

\author[A Chakraborty et al.]{
Arnab Chakraborty,$^{1}$\thanks{E-mail: phd1601121009@iiti.ac.in}
Abhirup Datta,$^{1,2}$
Samir Choudhuri,$^{3}$
Nirupam Roy $^{4}$
\newauthor
Huib Intema,$^{5}$
Madhurima Choudhury,$^{1}$
Kanan K. Datta,$^{6}$
Srijita Pal,$^{7}$
\newauthor
Somnath Bharadwaj, $^{7}$
Prasun Dutta, $^{8}$
Tirthankar Roy Choudhury $^{3}$ 
\\
$^{1}$Indian Institute Of Technology, Indore, India\\
$^{2}$University Of Colorado,Boulder,USA\\
$^{3}$National Centre For Radio Astrophysics,Post bag 3,Ganeshkhind,Pune 411007,India\\
$^{4}$Department of Physics , Indian Institute of Science,Bangalore 560012,India\\
$^{5}$Leiden Observatory, Leiden University, Niels Bohwerg 2, NL-2333CA, Leiden, The Netherlands\\
$^{6}$Department of Physics, Presidency University, 86/1 College Street, Kolkata-700073, India\\
$^{7}$Department of Physics $\&$ Centre for Theoretical Studies,IIT Kharagpur,Kharagpur 721302,India\\
$^{8}$Department of Physics, IIT (BHU) Varanasi, 221005 India
}
\date{Accepted XXX. Received YYY; in original form ZZZ}

\pubyear{2015}

\begin{document}
\label{firstpage}
\pagerange{\pageref{firstpage}--\pageref{lastpage}}
\maketitle

\begin{abstract}
In this first paper of the series, we present initial results of newly upgraded Giant Meterwave Radio Telescope (uGMRT) observation of European Large-Area ISO Survey-North 1 (ELAIS-N1)  at 325 MHz with 32 MHz bandwidth. Precise measurement of fluctuations in Galactic and extragalactic foreground emission as a function of frequency as well as angular scale is necessary for detecting redshifted 21-cm signal of neutral hydrogen  from Cosmic Dawn, Epoch of Reionization (EoR) and post-reionization epoch. Here, for the first time we have statistically quantified the Galactic and extragalactic foreground sources in the ELAIS-N1 field in the form of angular power spectrum using the newly developed Tapered Gridded Estimator (TGE). We have calibrated the data with and without direction-dependent calibration techniques. We have demonstrated the effectiveness of TGE against the direction dependent effects by using higher tapering of field of view (FoV). We have found that diffuse Galactic synchrotron emission (DGSE) dominates the sky, after point source subtraction, across the angular multipole range  $ 1115 \leqslant \mathcal{\ell} \leqslant 5083 $  and $ 1565 \leqslant \mathcal{\ell} \leqslant 4754 $ for direction-dependent and -independent calibrated visibilities respectively. The statistical fluctuations in DGSE has been quantified as  a power law of the form $\mathcal{C}_{\mathcal{\ell}}= A \mathcal{\ell}^{-\beta} $.  The best fitted values of  (A, $\beta$) are ($ 62 \pm 6$ $mK^{2}$, $2.55 \pm 0.3 $) and  ($ 48 \pm 4$ $mK^{2}$, $2.28 \pm 0.4 $ ) for the two different calibration approaches. For both the cases, the power law index is consistent with the previous measurements of DGSE in other parts of sky. 
\end{abstract}




\section{Introduction}

Observations of the  redshifted 21-cm ``spin-flip'' transition \citep{Field1958} of the neutral hydrogen (HI)  is considered as a promising probe for physical conditions in the early universe (For review: \citealt{Furlanetto2006,Morales2010}). Observations of Gunn-Peterson trough in  quasar absorption spectra \citep{Fan2002,Fan2006,Mortlock2011} and Thompson optical depth as measured from  from CMB temperature and polarization angular spectra  \citep{Planck Collaboration2018} together imply that universe was reionized (Epoch of Reionization) over a redshift range {\bf ($6<  \textit{z} <15$)}.   Studying the early universe through the redshifted 21-cm signal will be the first hint  to understand the nature of the first stars, galaxies and black holes and the  evolution of large-scale structures in the universe  \citep{Madau1997, Bharadwaj2001a, Fan2006}. 
The measurement of   HI 21-cm power spectrum along with tomographic imaging of the IGM using large interferometric arrays  holds the greatest potential to observe the redshifted HI 21-cm line \citep{Bharadwaj2001,Bharadwaj2005,Morales2004,Zaldarriag2004}.  Several upcoming and ongoing projects such as Donald C.Backer Precision Array to Probe the Epoch of Reionization (PAPER, \citealt{Parsons2010,Kerrigan2018}), the Low Frequency Array(LOFAR, \citealt{vanHaarlem2013}), the Murchison Wide -field Array (MWA, \citealt{Li2018}), the Square Kilometer Array (SKA1 LOW, \citealt{Koopmans2015}) and the Hydrogen Epoch of Reionization Array (HERA, \citealt{DeBoer2017}) will have needed sensitivity to  measure redshifted HI 21-cm power spectrum. However, we may need the complete SKA to do a successful tomographic imaging of IGM.

In addition, statistical detection of intensity fluctuations in post-reionization 21-cm signal  ($  \textit{z}  \lesssim 6$ or $\nu \gtrsim 200 $~MHz), using  intensity mapping experiment,  provides a  unique tool for precision cosmology. Mapping of 21-cm intensity fluctuations in post-reionization era can  quantify the large scale HI power spectrum, source clustering, etc \citep{Bharadwaj2001a,Bharadwaj2003,Bharadwaj2005,Wyithe2008}.  BAOBAB \citep{Pober2013a}, BINGO \citep{Battye2012}, CHIME \citep{Bandura2014}, the Tianlai project \citep{Chen2016}, HIRAX \citep{Newburgh2016}, SKA1-MID \citep{Bull2015}  will measure Baryon Acooustic Oscillations (BAO) over a redshift range  $ {\bf \textit{z}}$ $\sim 0.5 - 2.5 $, which can be used as a standard ruler to constrain Dark Energy equation of state. Efforts are also ongoing to make a $ \sim 5\sigma$ detection of amplitude of power spectrum $ A_{HI}$   around {\bf \textit{z}} $\sim 3.35 $ using  OWFA \citep{Subrahmanya2017} .

The expected brightness temperature of redshifted HI-signal from the EoR and post-reionization epoch  is many orders of magnitude fainter than the radio emissions from  different Galactic and extragalactic foregrounds \citep{Bharadwaj2005,Zaldarriag2004}.  The challenges are nearly identical for both EoR and post-reionization experiments. So the knowledge of  foregrounds at post-reionization epoch can also help us to understand the intricacies involved in  detection of the HI signal coming from EoR. Accuracy of  extraction of the  cosmological signal strongly depends on the ability to characterize and remove the foregrounds from observational data sets at the frequency of redshifted HI 21-cm line. Depending upon sensitivity we can identify individual sources and remove them from the image down to a certain flux level. But the effect of residual  sources to the power spectrum could overwhelm the cosmological signal \citep{Di Matteo2002,Datta2009}. The foreground sources are diffuse Galactic synchrotron emission from our galaxy (DGSE) \citep{Shaver1999}, free-free emission from ionizing haloes \citep{Oh2003}, faint radio-loud quasars \citep{Di Matteo2002}, synchrotron emission from low-redshift galaxy clusters \citep{Di Matteo2004},etc. 

Previous studies have shown that  foreground spectra from astrophysical sources  are generally smooth and correlated over a frequency separation of $\bigtriangleup \nu \sim 1$ MHz whereas the HI-signal decorrelates rapidly over such frequency separation \citep{Bharadwaj2001,Bharadwaj2005}. This property allows us to separate the cosmological signal from the foregrounds \citep{Ghosh2011}. There are mainly three different techniques used to deal with foregrounds- foreground avoidance  \citep{Datta2010,Datta2010a,Trott2012,Pober2013},  foreground suppression \citep{Chapman2013,Choudhuri2016} and  foreground removal \citep{Datta2009,Chapman2016}.

  The   basic concept of foreground removal technique is to model each foreground components precisely and subtract that model from the data set. Modelling bright point sources residing at the edge of FoV is difficult because the primary beam becomes asymmetric and highly time and frequency dependent at outer part of the FoV. It is possible to suppress the effects of bright sources at the edge of FoV by tapering the sky response using  Tapered Gridded Estimator (TGE) \citep{Choudhuri2014,Choudhuri2016} (See: Sec. \ref{TGE_theory}).

Characterizing  foregrounds to the best possible extent with low frequency observations  is essential to construct accurate model of foregrounds.  Using these sensitive observations, we can also learn about properties of extragalactic point sources and DGSE, which  apart from being  two main foreground components are scientifically interesting in its own right. In addition, the knowledge of fluctuations in  the Galactic synchrotron emission can be used to probe  structures and magnetic field in interstellar medium (ISM) of the Milky Way \citep{Waelkens2009,Lazarian2012,Iacobelli2013}.

 In this paper, we have studied fluctuations in foregrounds of  European Large-Area ISO Survey-North 1 (ELAIS-N1) with uGMRT at 325 MHz. ELAIS-N1 has been previously studied at other frequencies \citep{Garn2008,Sirothia2009,Jelic2014,Taylor2016}. The field lies at high galactic latitude ($b=+44.48^{\circ}$), therefore the contribution of  Galactic synchrotron emission to foregrounds is relatively small for this patch of sky. This helps us to quantify extragalactic foreground sources  with the main motivation to detect redshifted HI signal from  post-EoR.  This is going to be the first among a series of papers from the deep (25 hrs) observation of ELAIS-N1 field at this frequency.  We will systematically study this field with final motivation to get upper limit on post-EoR signal. As a first step, here  we present the detailed analysis of the GSB data set (32 MHz bandwidth)  and effectiveness of TGE  to estimate angular power spectrum of  point sources and DGSE.   We have also studied the effect of  different  calibration techniques in estimation of power spectrum of DGSE.  In forthcoming paper we will present the detailed analysis of  the GMRT Wideband Backend  (GWB) data set (200 MHz bandwidth),  source catalog, differential source counts, cross- correlation between sources detected in other  wavelengths, characterization of foreground with respect to full bandwidth (200 MHz), etc. 
    
This paper is structured as follows. In Sect. \ref{Low_frequency_obs} a brief summary of existing low frequency observations for different fields are mentioned. We  describe the uGMRT observations of ELAIS-N1  in  Sect. \ref{ uGMRT observation}. The details of RFI mitigation and direction-independent calibration and imaging are mentioned in Sect. \ref{Data_reduction}. For direction-dependent calibration  basic work methodology of SPAM is given in Sect. \ref{SPAM_theory}. We have applied  Tapered Gridded Estimator (TGE) \citep{Choudhuri2014,Choudhuri2016} to both direction-independent and direction-dependent calibrated visibilities to determine the effect of different calibration techniques on estimation of angular power spectrum ($\mathcal{C}_{\mathcal{\ell}} $). A brief theory of TGE and  results are presented in Sect. \ref{TGE_theory}. Finally, Sect. \ref{conclusion} summarizes and concludes this work.

\section{Low Frequency Radio Universe - Physics and Observations}
\label{Low_frequency_obs}
The gyration of cosmic ray electrons in the magnetic field of our Galaxy is the main source of synchrotron radiation. The energy spectrum and density of cosmic ray electrons and also the magnetic field strength vary across the Galaxy. Therefore, observed synchrotron radiation will depend on frequency of observation as well as on the patch of sky we are observing through radio interferometer. Radio observations at $\nu \leq 1.4GHz$ provide the clearest picture of the Galactic synchrotron morphology, since at these frequencies the
diffuse  non-thermal radiation  clearly  dominates  over  all  other emissions  outside the Galactic plane. There are several observations covering different  regions of sky spanning a wide range of frequencies to characterize DGSE. There is an all-sky map of Galactic synchrotron radiation by \citet{Haslam1982} at 408 MHz. A map of DGSE at 1420 MHz have presented by \citet{Reich1982} and \citet{Reich1988}. \citet{Giardino2001} have shown using Rhodes survey at 2.3 GHz that angular power spectrum ($\mathcal{C}_{\mathcal{\ell}}$) of DGSE behaves like a power law, 
\begin{equation}
\mathcal{C}_{\mathcal{\ell}}= A \times (1000/ \mathcal{\ell})^{\beta}
\end{equation}
where the power law index $\beta=2.43 $ in the $\mathcal{\ell}$ range  $ 2 \leqslant \mathcal{\ell} \leqslant 100 $. \citet{Giardino2002} have found $\beta=2.37 $ in the $\mathcal{\ell}$ range $ 40 \leqslant \mathcal{\ell} \leqslant 250 $ for the Parkes survey at 2.4 GHz. \citet{Bernardi2009} have analyzed 150 MHz WSRT observation to characterize the fluctuations in DGSE and found that $ A=253$ $mK^{2}$ and $\beta=2.2$ for $\mathcal{\ell}  \leqslant 900 $. \citet{Ghosh2012} have reported  $A=513$ $mK^{2}$ and $\beta=2.34$ in the  $\mathcal{\ell}$ range $ 253 \leqslant \mathcal{\ell} \leqslant 800 $ using 150 MHz GMRT observations. \citet{Iacobelli2013}  have reported using LOFAR observation at  160 MHZ  that fluctuations in DGSE ($\mathcal{C}_{\mathcal{\ell}}$) approximately follows a power law with a slope $\beta \approx 1.8 $ up to $\mathcal{\ell}$ = 1300. 

\citet{Ali2008} have studied the foregrounds on sub-degree angular scales with GMRT observation at 150 MHz. They have used the correlations among  measured visibilities to directly determine the multi-frequency angular power spectrum $\mathcal{C}_{\mathcal{\ell}}$ $(\bigtriangleup \nu)$ (MAPS, \citet{Datta2007}). They have found that the measured
$\mathcal{C}_{\mathcal{\ell}}$ $(\bigtriangleup \nu)$ before point source subtraction has a value around $10^{4}$ $mK^{2}$. This is seven order of magnitude stronger than the expected redshifted HI-signal. 

\citet{La Porta2008} have calcultaed  that angular power spectrum (APS) of DGSE as a function of Galactic latitudes by considering various cuts in the sky. They have found that APS  is best fitted with a power law and the power law index lies between [2.6 - 3] for different Galactic latitudes. 

\citet{Choudhuri2017} have analyzed two different fields of TIFR GMRT Sky Survey (TGSS) at 150 MHz near the Galactic plane ($9^{\circ} ,+10^{\circ}$) and ($15^{\circ} ,-11^{\circ}$) to characterize the statistical properties of DGSE . They have found that the measured total intensity of angular power spectrum shows a power law behavior in the $\mathcal{\ell}$ range $ 240 \leqslant \mathcal{\ell} \leqslant 580$ and $ 240 \leqslant \mathcal{\ell} \leqslant 440 $  and the best fitted values of ($A,\beta $) are ( 356,2.8 ) and (54, 2.2) for two different fields  respectively. 

The outcome of all these analysis is that angular power spectrum of synchrotron radiation over large portion of the sky can be modelled as a power law of the form $\mathcal{C}_{\mathcal{\ell}}= A (1000/\mathcal{\ell})^{\beta} $ with $\beta \sim [1.5,3.0]$ for $\mathcal{\ell} \leq 1300 $, corresponding to a angular scale $\theta \geq 0.2^{\circ}$. This general result does not include the complexity of the angular power spectrum of synchrotron emission whose parameters are expected to change with frequency and sky direction.
\begin{table}
	\centering
	\caption{Observation Summary for GWB and GSB}
	\begin{tabular}{lccr} 
		
		\hline
            &   GWB & GSB\\
       \hline
		Working antennas & 28 & 28 \\
		Central Frequency & 400 MHz & 325 MHz \\
		Bandwidth & 200 MHz & 32MHz\\
		Visibility integration time & 2 sec & 8 sec\\
        Number of Channels & 8192 & 512  \\
		Total Observation time & 25 hours &25 hours \\
        Frequency resolution & 24 KHz & 65 KHz \\
        
        \hline
	\end{tabular}
\label{observation}    
\end{table}
   
\begin{table}
\centering
\caption{Detail of Calibrators of this observation}
\begin{tabular}{lccr}
\hline
        Flux Calibrator \\
		Source & 3C286 \\
		Flux Density & 23 Jy \\
        Source & 3C48 \\
        Flux Density &  42Jy\\
		Scale & Scaife-Heald \\
        \hline
		
         Phase Calibrator \\
		Source & J1549+506 \\
		
		Flux Density & 0.3 Jy \\
        \hline
		
        Target Field \\
		Source &  ELAIS N1 \\
		time   &   14 hours           \\
		\hline
\end{tabular}
\label{calibrator}    
\end{table}

\section{\lowercase{u}GMRT Observation}
\label{ uGMRT observation}
The Giant Metrewave Radio Telescope (GMRT) \citep{Swarup1991} is one 
of  the  largest  and  most  sensitive  fully  operational  low-frequency  radio  telescopes  in  the  world  today.  The  array  
configuration  of  30  antennae  (each  of  45m  diameter)  
spanning  over  25 km  provides  a  total  collecting  area  of  
about  30,000  Sq.m  at  meter  wavelengths,  with  a  fairly  
good angular resolution ($\sim$arcsec). Out of the 30 antennae, 14 antennae are randomly distributed in a Central square which is approximately 1.1 Km $\times$ 1.1 Km in extent. The rest of the antennae lie along three nearly 14 Km long arms in an approximately `Y' shaped configuration. Recently GMRT has been upgraded to uGMRT with some extra features, such as: (i) huge frequency coverage, from 120-1500 MHz; (ii) maximum bandwidth available is 400 MHz instead of 32 MHz bandwidth of original GMRT design; (iii) digital backend correlator catering to 400 MHz bandwidth; (iv) improved  receiver  systems  with higher $G/T_{sys}$ and  better  dynamic  range \citep{Gupta2017}.

\begin{figure*}
\centering
\includegraphics[width=2\columnwidth]{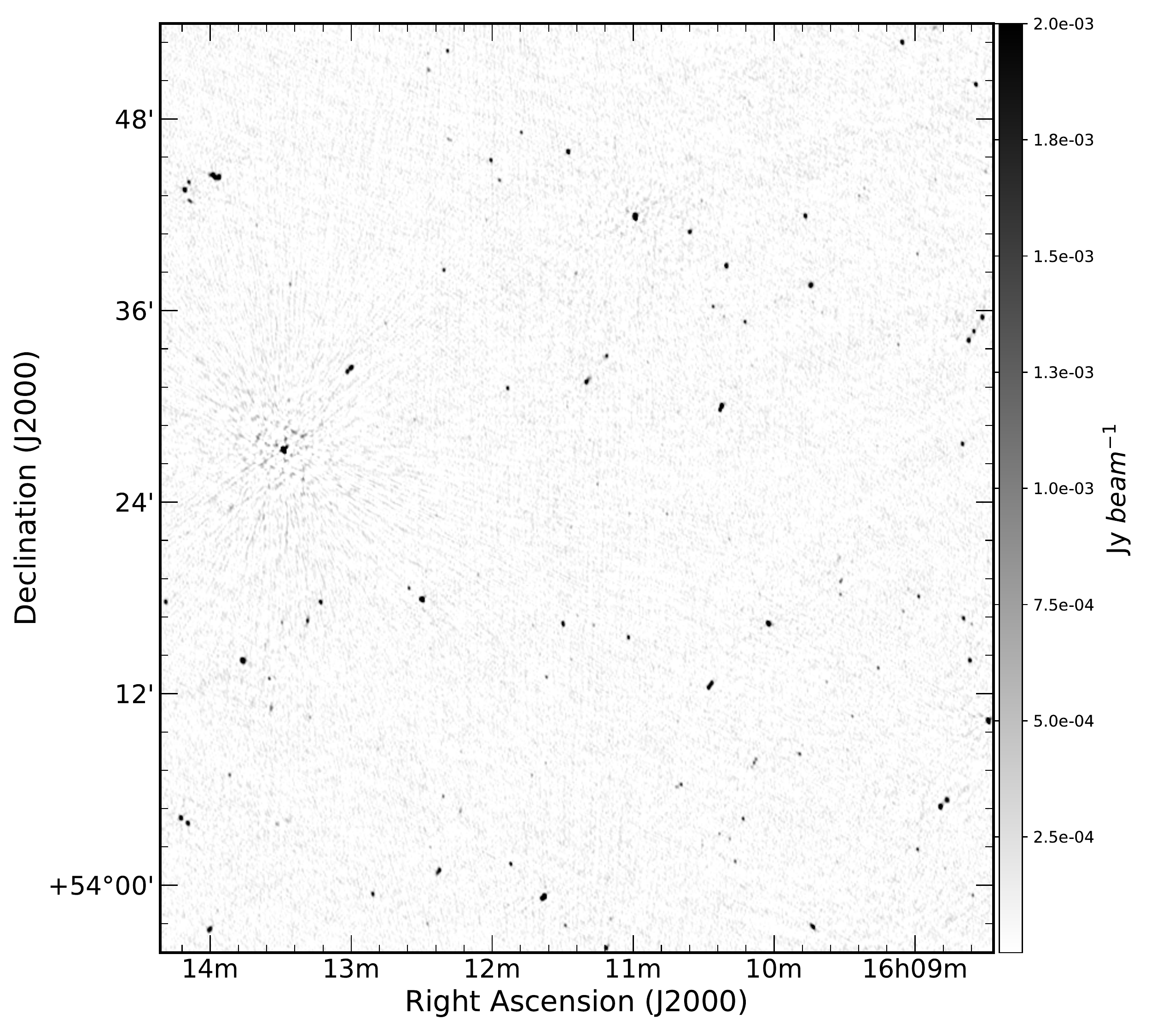}

\caption{The above uGMRT image is total intensity map of ELIAS N1 at 325 MHz (bandwidth 32 MHz) after analyzing the data with CASA where only direction independent calibration has been performed. The rms achieved is 80 $\mu$Jy and beam size is $ 11\arcsec \times 6\arcsec$ and the dynamic range is $\sim 4000 $. }
\label{fig.CASA}
\end{figure*}

\begin{table*} 
	\begin{center}
	\caption{Imaging Summary}
	\begin{tabular}{lccr}
    \hline
            &   Direction Independent(CASA) & Direction Dependent (SPAM)\\
       \hline
       
		\hline
        Image size & 4096$\times$4096 &3582$\times$3582  \\
        \hline
		pixel size  & 2.0$\arcsec$ $\times$ 2.0$\arcsec$ & 2.0$\arcsec$ $\times$ 2.0$\arcsec$ \\
        \hline
		Number of  wprojection planes & 256  & 256 \\
        \hline
		Off-source noise &  80$\mu$Jy/Beam  &  40$\mu$Jy/Beam \\
        \hline
		Dynamic range( Peak/noise) &  4000 & 10000\\
        \hline
        Flux Density (max,min) & (350 $\mu$Jy , -12 mJy) & (390 mJy , -6 mJy) \\
        \hline
        Synthesized Beam      & 11$\arcsec$ $\times$ 6$\arcsec$ & 11$\arcsec$ $\times$ 8$\arcsec$\\
        \hline
        \hline
	\end{tabular}
     \end{center}
\label{imaging}	    
\end{table*}

We carried out deep observation of the ELAIS-N1 field ($\alpha_{2000}=16^{h}10^{m}1^{s} ,\delta_{2000}= 54^{\circ}30'36\arcsec$) with the uGMRT in GTAC (GMRT Time Allocation Committee) cycle 32 during May 2017 for 25 hours over four days. The ELAIS-N1 field lies at high Galactic latitudes ($\mathcal{\ell}=86.95^{\circ} ,b=+44.48^{\circ}$) and was up at night time during the GTAC cycle 32 and the field contains relatively few bright sources.
The observation was carried out at night for all days to minimize the Radio Frequency Interference (RFI) from man made sources. Further the ionosphere is considerably more stable at night. We have conducted  the observation for long time (25 hrs) to achieve high dynamic range and to get adequately sampled visibilities for further statistical analysis. 
For each observing session,  we have  observed  a flux calibrator 3C286 in the beginning and 3C48 at the end of the observation run. We have observed a phase calibrator J1549+506 (near the target field) in every 25 minutes to correct for  the temporal variations in the system gain. 
 
   The observation summary along with details of the calibrators  are  presented in Table \ref{observation} and Table \ref{calibrator}. The observational setup was a total 512 frequency channels spanning 32 MHz bandwidth centered at 325 MHz using GMRT Software Backend (GSB). The time and frequency resolution of the observation are 8s and 65 KHz respectively.

 \section{Data Analysis}
 \label{Data_reduction}

\subsection{ RFI mitigation} Radio frequency interference (RFI)  limits the sensitivity of radio observations by increasing the system noise and corrupting the calibration solutions. It also restricts the available frequency bandwidth. The effect is particularly strong at frequencies below 600 MHz at GMRT.
       
There is  ringing  across frequency channels that neighbor the strong, usually narrow, RFI. This phenomenon usually  known as Gibbs ringing. To mitigate this ringing we have employed Hanning-smoothing algorithm in CASA. Hanning-smoothing applies a triangle as a smoothing kernel across the spectral axis  which diminishes the ringing and also reduces the number of channels that may look bad by flagging them. As a result  spectral resolution has been decreased. 
After that we have used  an autoflag algorithm, \textit{RFLAG}, for RFI excision. In order to get the best possible result from  \textit{RFLAG}, we have first solved for an initial set of antenna-based phases over a narrow range of channels and made an average bandpass over the entire observing session using phase calibrator. However, for final bandpass  calibration we have used 3C286, which gives higher signal to noise ratio in the bandpass.  bf   We have applied \textit{RFLAG} to the bandpass corrected data, where data is iterated through in segments of time and local rms and median rms of real and imaginary part of the visibilities across channels as well as across a sliding time window has been calculated. Deviation of local rms from this median value is being calculated. If local rms is larger than  5 times the median value of deviation, then the data was flagged. The bulk of flagging is done using \textit{RFLAG} on all data uniformly for direction-independent and -dependent calibration. Rest is minor iterative flagging during calibration steps.

\subsection{Direction-Independent approach}
 After flagging of spurious signal present in the data set, we have done direction-independent calibration using CASA.  We have calibrated individual night's data separately.  The calibration is done with exactly same parameters for different night's data sets. During imaging we have used all of them  to make a combined continuum image.

{\bf Calibration :}  We have used 3C286 as flux density and bandpass calibrator. We have used \citet{Scaife2012} model to set the flux value of 3C286 and 3C48 using \textit{SETJY} task in CASA. Using CASA task \textit{BANDPASS}, we have first done bandpass calibration  to account for gain variation as a function of frequency for 3C286. Then we have calculated  gain and phase variations as a function of time  on a 16s time scale, using \textit{GAINCAL}, for all the calibrators ,i.e, for flux calibrator (3C286), phase calibrator (J1549+506) and for our last scan of the calibrator 3C48. While we know the flux density of our primary calibrator (3C286), the model assumed for the secondary calibrator (J1549+506) was a point source of 1 Jy located at the phase center. 
We have used the 3C286 to determine the system response to a source of known flux density and used this to find out the true flux density of J1549+506. 

 We have applied flux density, bandpass, gain and phase calibration solutions  from phase calibrator (J1549+506) to the target field ELIAS-N1, since  it is near to the target field. 
During calibration bad data were flagged in various stages. In antenna based solution, data for an antenna with large error was flagged. Some baselines were also flagged based on closure error. After calibration and RFI mitigation nearly 30 percents of on source data were flagged.   

{\bf Imaging and Self Calibration :} The field of view ($1.4^{\circ} \times 1.4^{\circ}$) is large for GMRT at 325 MHz. We have taken 256 $\textit{w}$projection planes in the CASA task \textit{CLEAN} with  gridmode=`widefield' to take into account the non-coplanar nature of the GMRT antenna distribution. We have used Briggs robust parameter -1 as this shifts slightly towards uniform weighting. This produces nearly Gaussian central PSF while suppressing the broad wings and  suppresses  the abundance of short baselines in GMRT observation. We have used  multi-scale multi-frequency (MS-MFS) deconvolution algorithm \citep{Rau2011} in CASA with  nterms=2 to account for the total intensity (Stokes I) as well as the  spectral term. Table \ref{imaging} contains a summary of the imaging details with all the relevant parameters which are mostly self-explanatory.

 We have carried out several rounds of self-calibration to  reduce the error from temporal variations in the system gain and spatial and temporal variations in the ionospheric properties.  For individual night's data, we have done phase only self -calibration on the target field for 4 rounds with gain-solutions 5min, 4min, 2min and again 2min respectively .
     The final continuum image is shown in the Figure \ref{fig.CASA}. We have created a large  image of size $2.3^{\circ} \times 2.3^{\circ} $ to include the bright sources at the edge of the FoV. Otherwise, side lobes of those sources will cause ripple along frequency direction and  distort the image. Here we present only the central zoomed in part of the image of size $1.2^{\circ}  \times 1.2^{\circ}$. The off-source rms of the image is 80$\mu$Jy and size of the synthesized beam is $ 11\arcsec \times 6\arcsec$. Note that there are localized imaging artefacts around bright sources due to residual phase errors which have not been corrected during   self-calibration.
     
\subsection{Direction-Dependent approach}
\label{SPAM_theory}
For direction-dependent calibration we have used a fully automated AIPS \citep{Greisen1998} based pipeline, Source peeling and atmospheric modeling (SPAM) \citep{Intema2016,Intema2014,Intema2009}. SPAM includes direction-dependent calibration , modeling and imaging for correcting mainly ionospheric dispersive delay. The pipeline uses ParselTongue interface \citep{Kettenis2006} to access AIPS task, files and tables from python.
\begin{figure*}
\includegraphics[width=2\columnwidth]{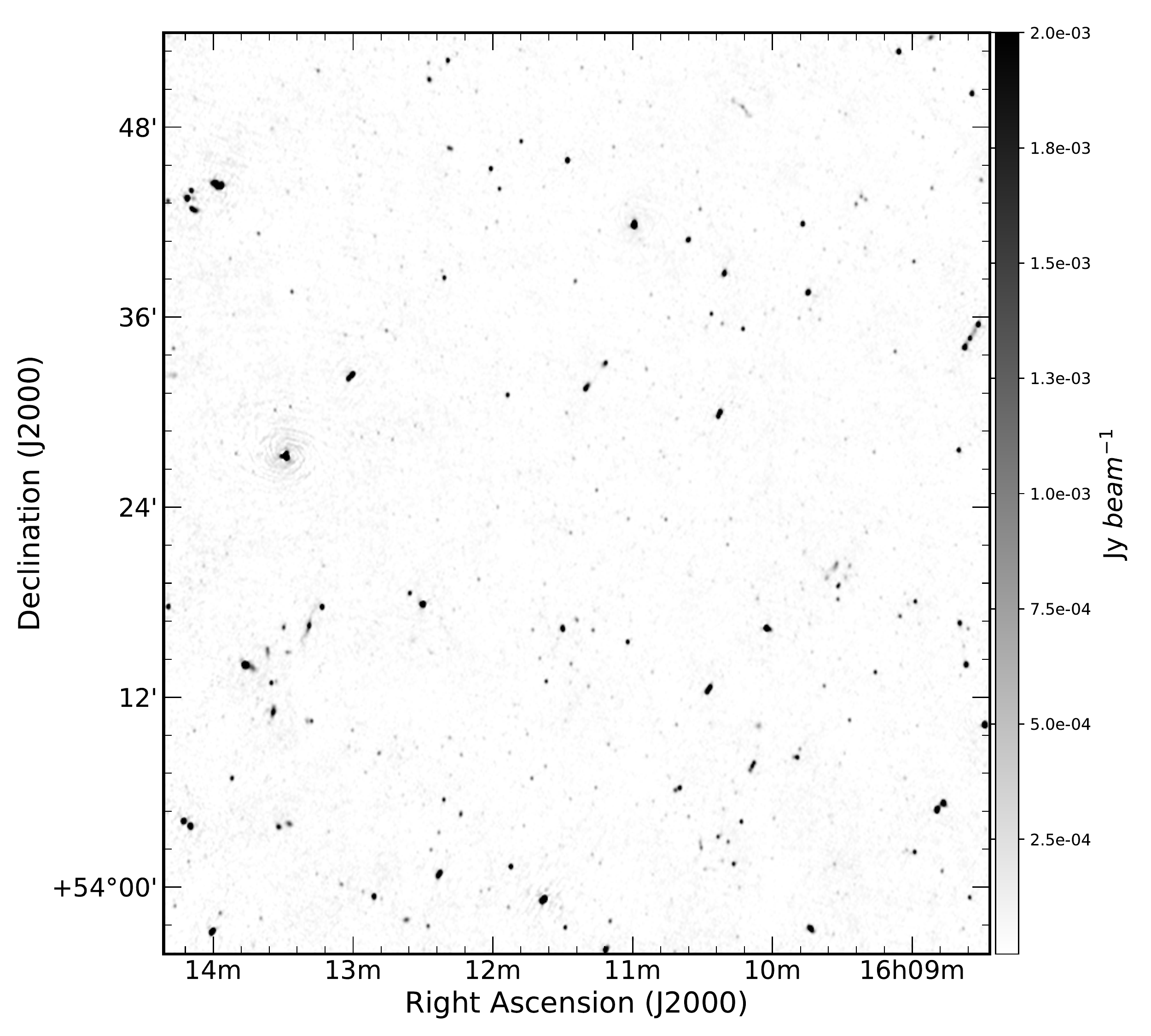}
\caption{The uGMRT 325 MHz total intensity image of ELAIS N1 after direction-dependent calibration has been performed with SPAM. The rms noise achieved is 40$\mu$Jy and synthesized beam is 11$\arcsec \times 8 \arcsec $ and the dynamic range is $\sim 10000$.}
\label{fig.SPAM}
\end{figure*}
SPAM consists of two parts: a pre-processing part that converts  raw data from individual observing session (LTA format) into pre-calibrated visibility data sets and a main pipeline part which  converts pre-calibrated visibility data into stokes I continuum image. 

{\bf Pre-Calibration :} In the pre-processing part, SPAM  computes good-quality instrumental calibration from the best available scan of one of the primary calibrators and apply these calibration to the data. The data for each day has been calibrated separately. Flux density of calibrators has been set following low-frequency flux models \citet{Scaife2012}. For each scan on each calibrator after initial flagging of RFI it determined time variable complex gain solution and time constant bandpass solution per antenna and per polarization. To reduce the data size and speed up the processing, LL and RR polarizations were combined as Stokes $\textit{I}$  and  data was averaged in frequency and time. The final number of channels after averaging is 42 of width 0.761 MHz  yielding an effective bandwidth of 32 MHz. The pipeline computes a weight factor, which is proportional to number of active antennae and inverse variance of the gain amplitude and the best calibrator scan is with the highest weight. It used  best scan of 3C286 and applied the calibration solution of this calibrator to the target data. After calibration, the $\textit{uv}$-data from all four days were combined.
  
\begin{figure*}
\centering
\begin{tabular}{cc}
 Direction Independent(CASA) & Direction Dependent (SPAM)\\
\includegraphics[width=2.5in,height=2.in]{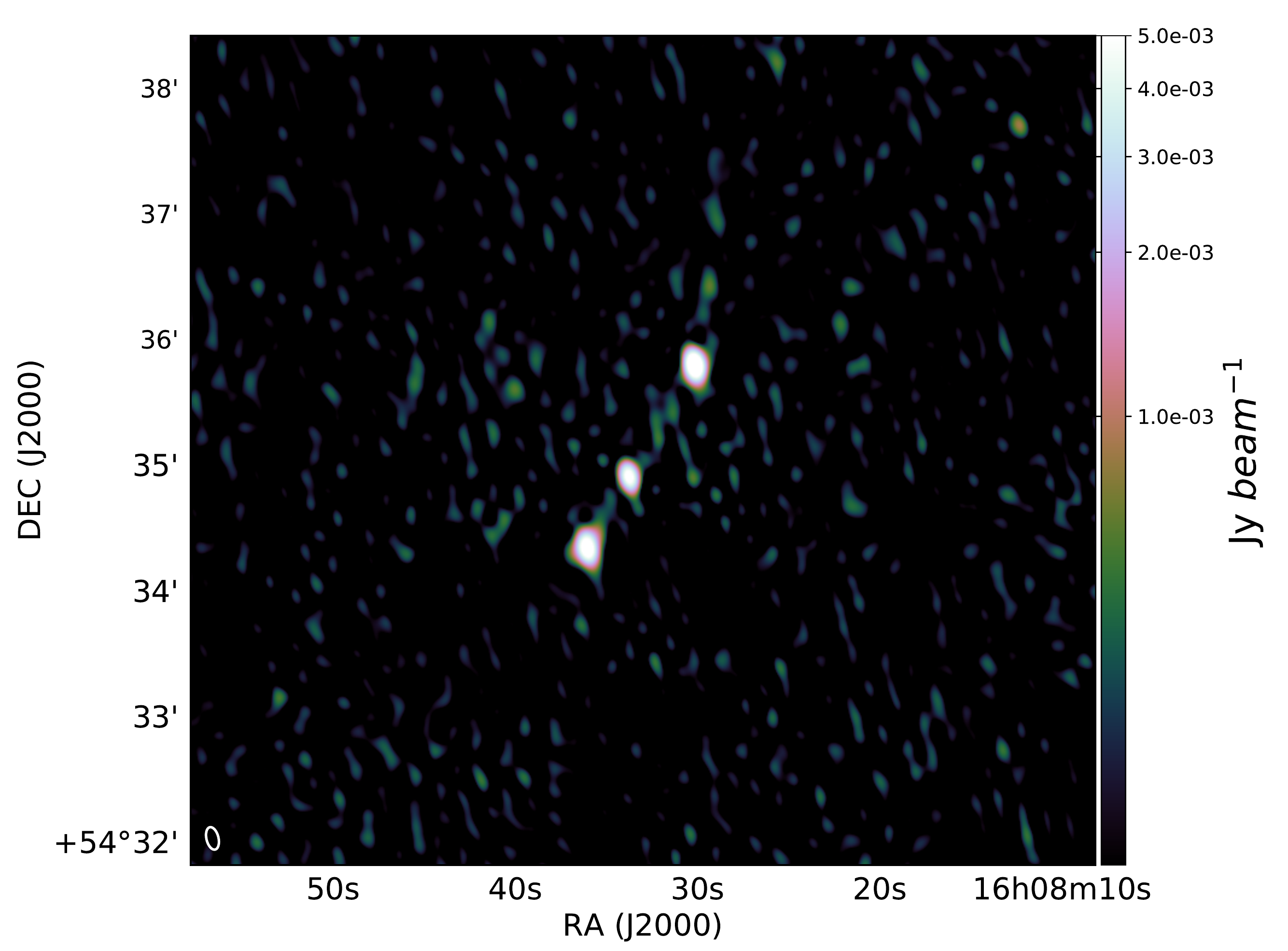}&
\includegraphics[width=2.5in,height=2.in]{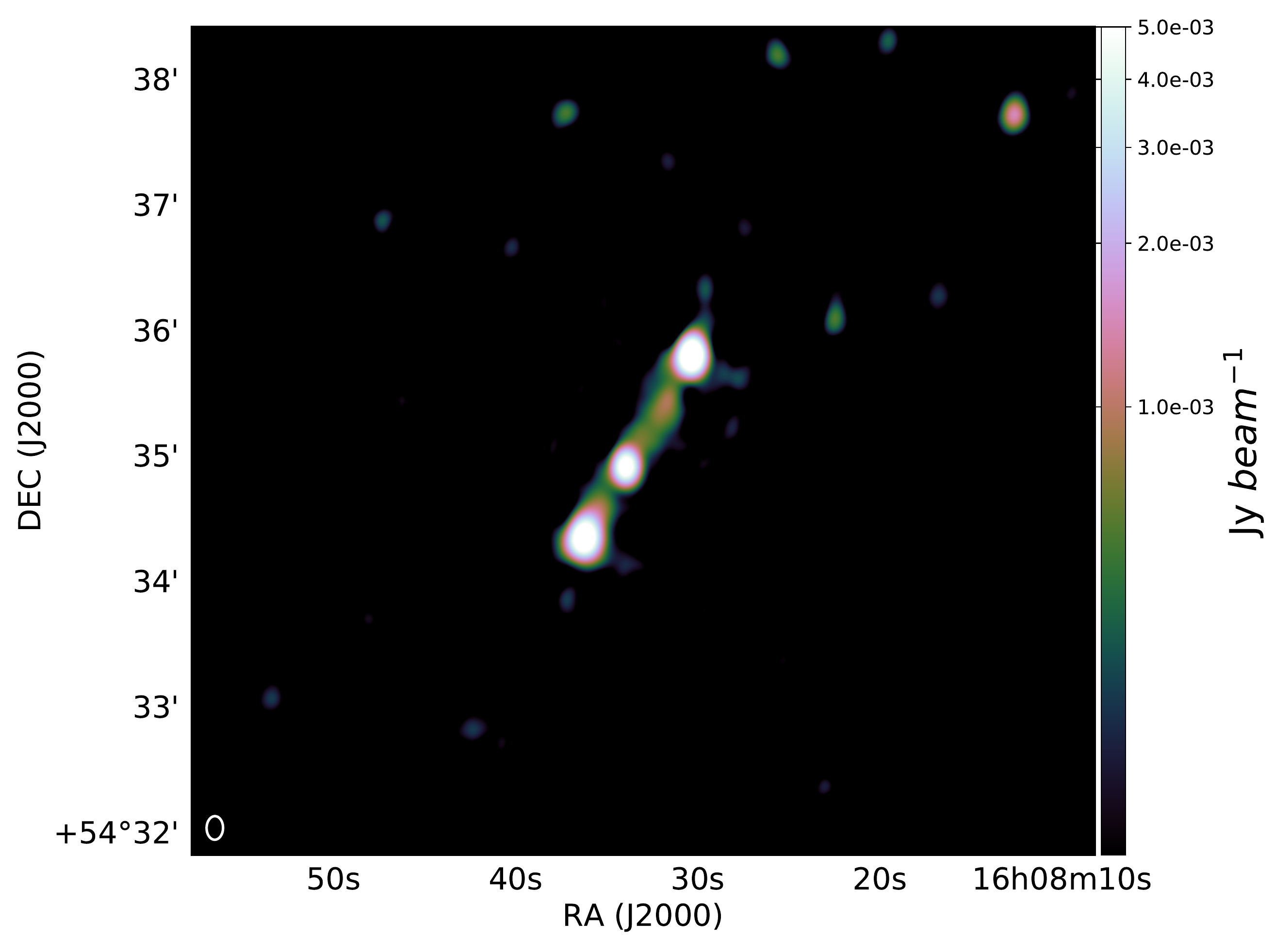}\\
\includegraphics[width=2.5in,height=2.in]{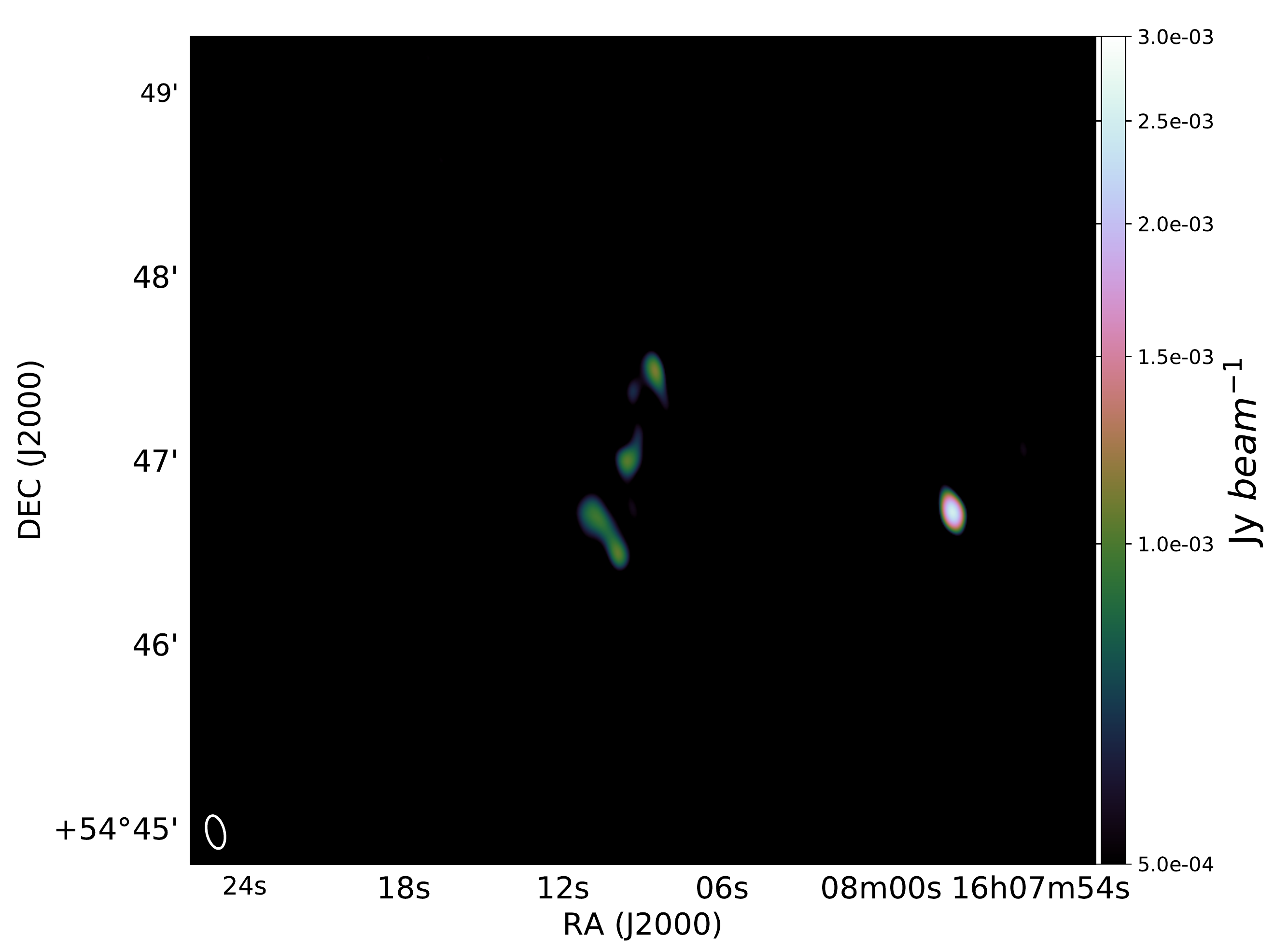}&
\includegraphics[width=2.5in,height=2.in]{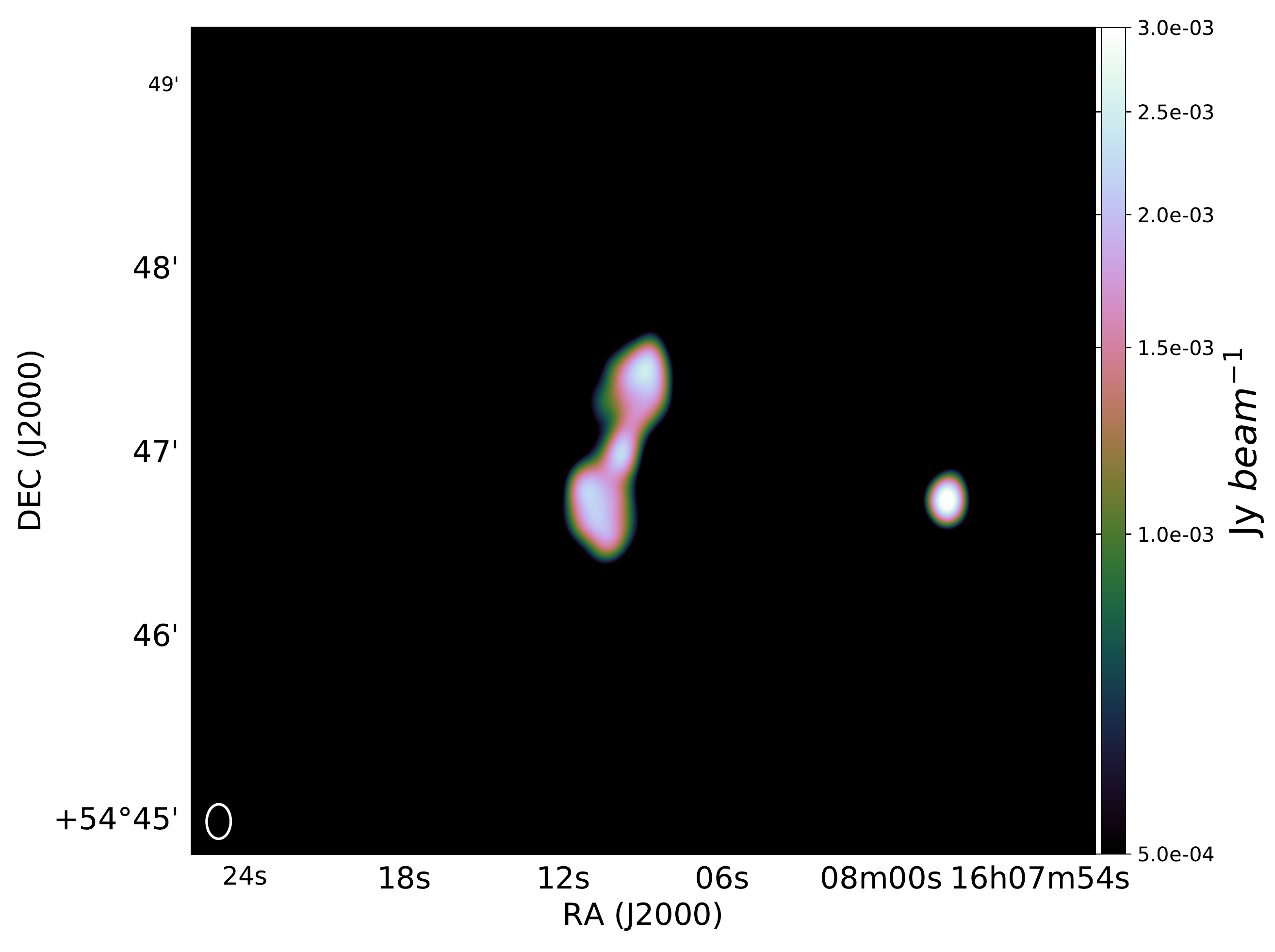}\\
\includegraphics[width=2.5in,height=2.in]{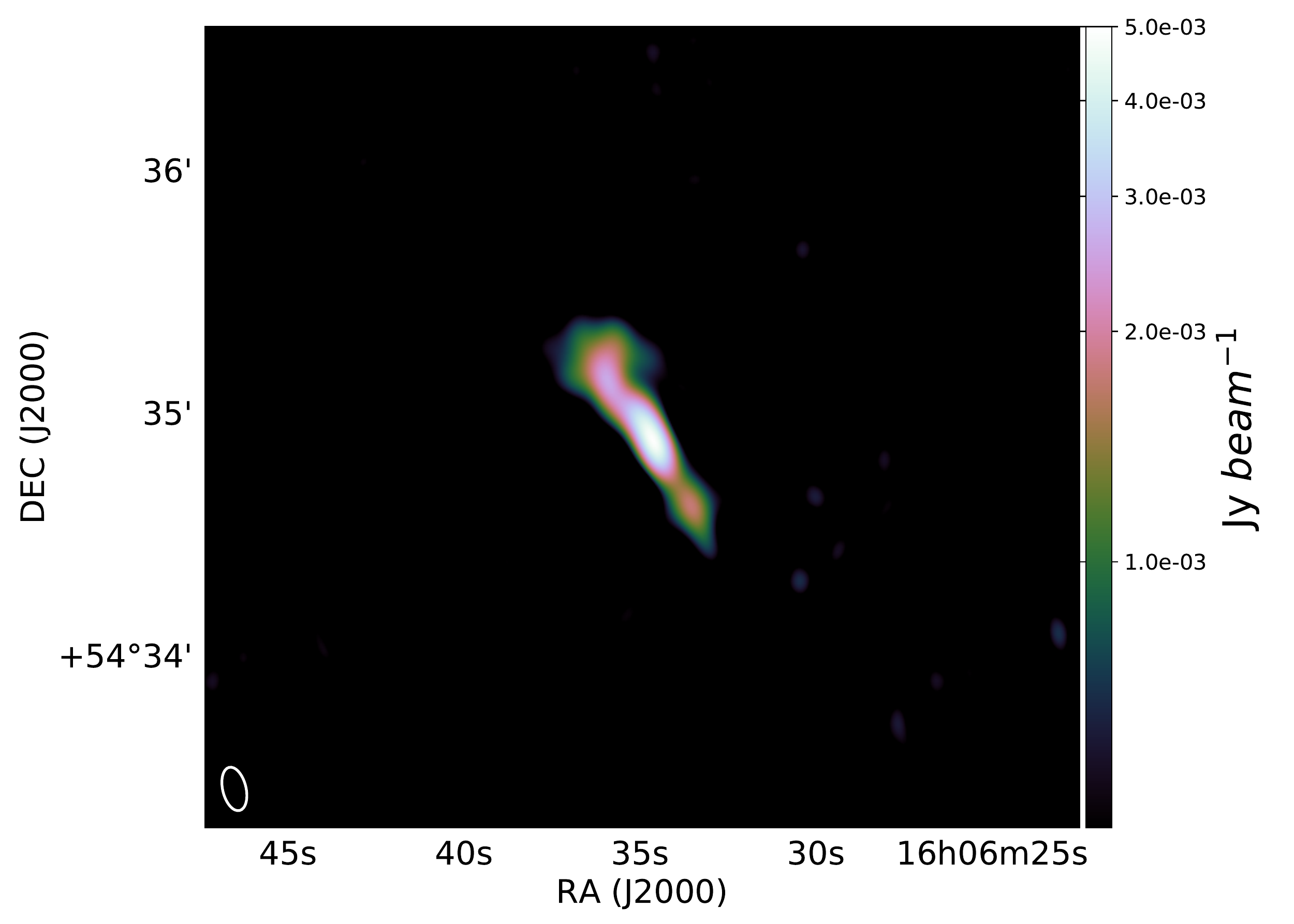}&
\includegraphics[width=2.5in,height=2.in]{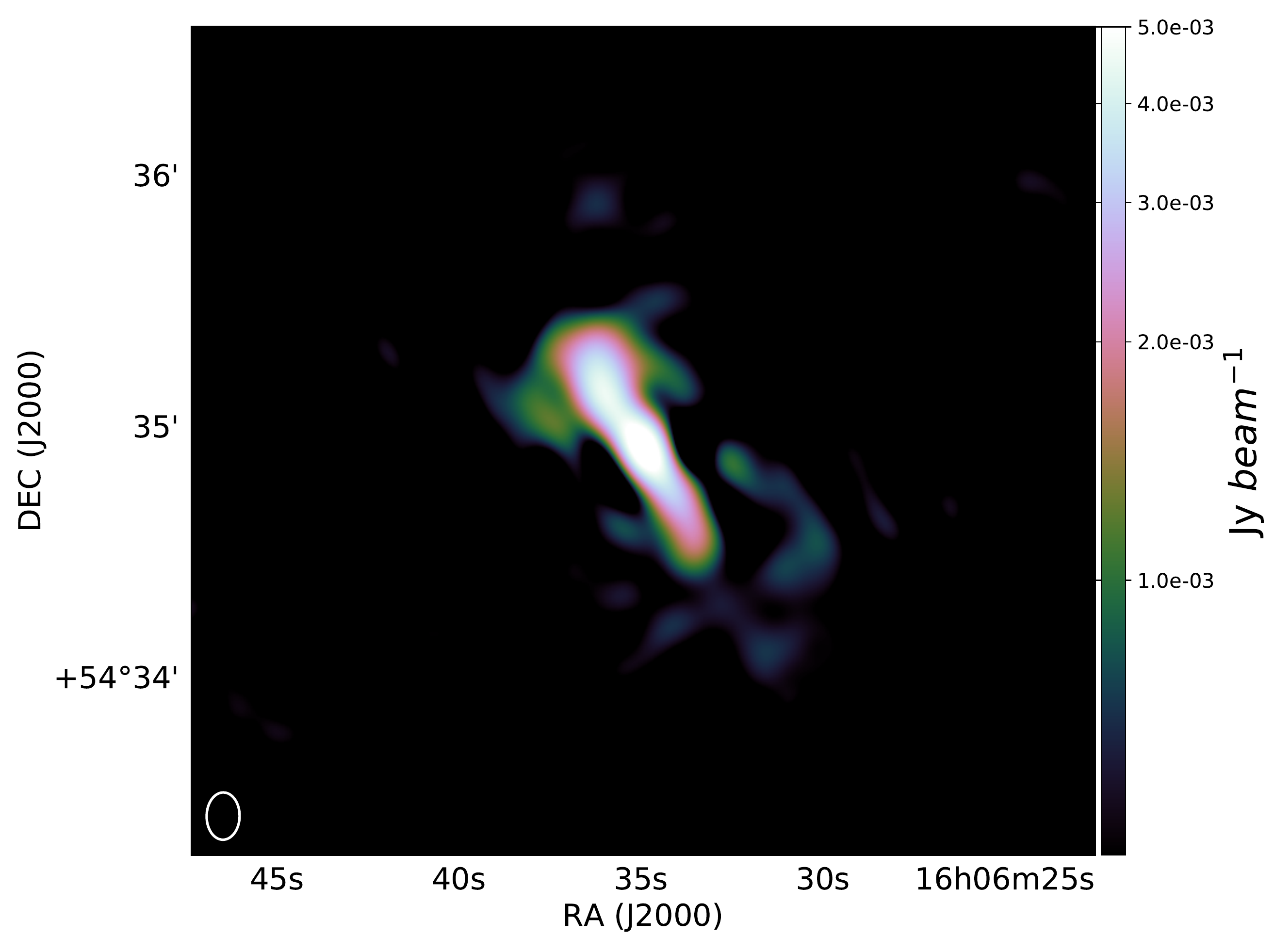}\\
\includegraphics[width=2.5in,height=2.in]{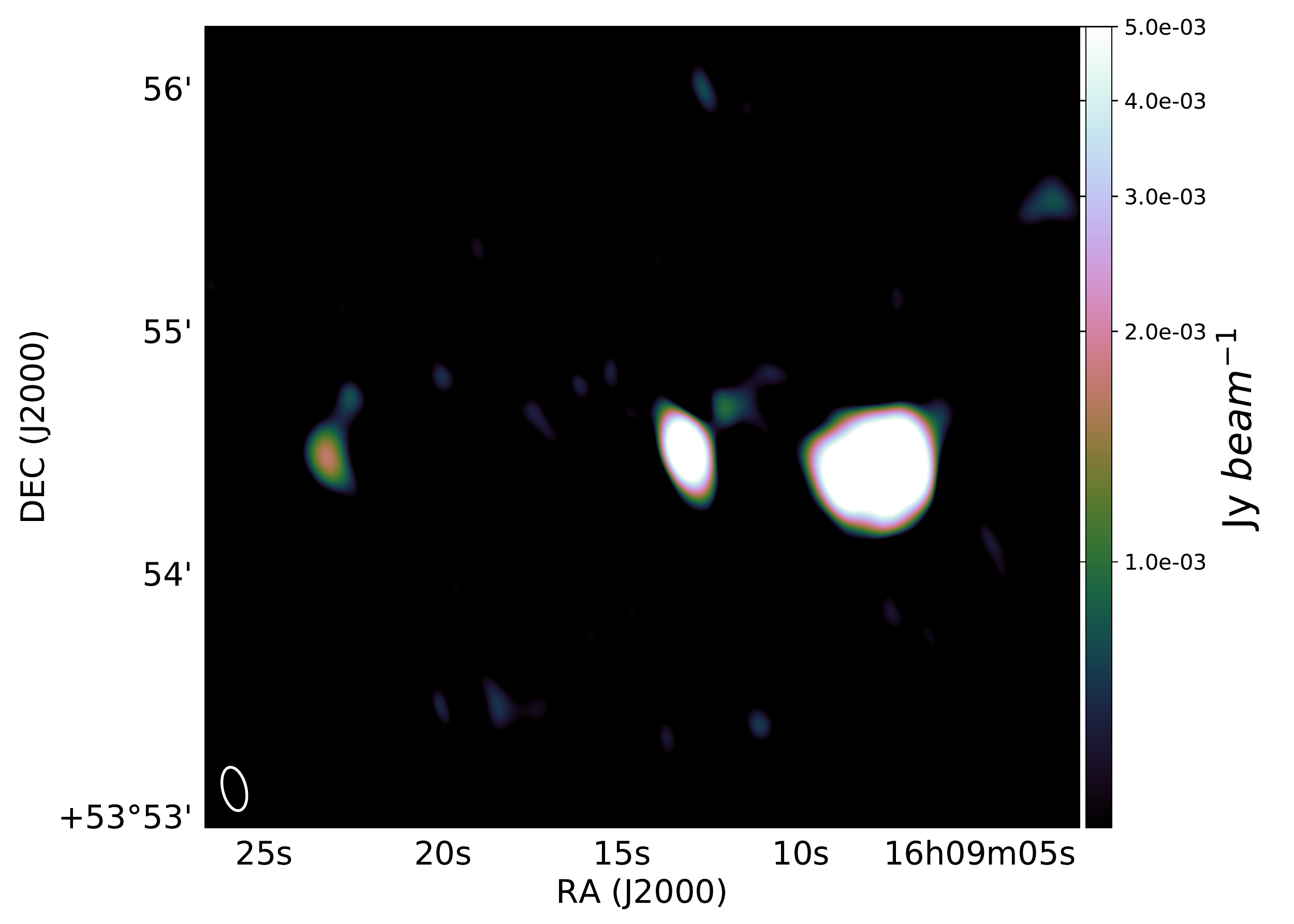}&
\includegraphics[width=2.5in,height=2.in]{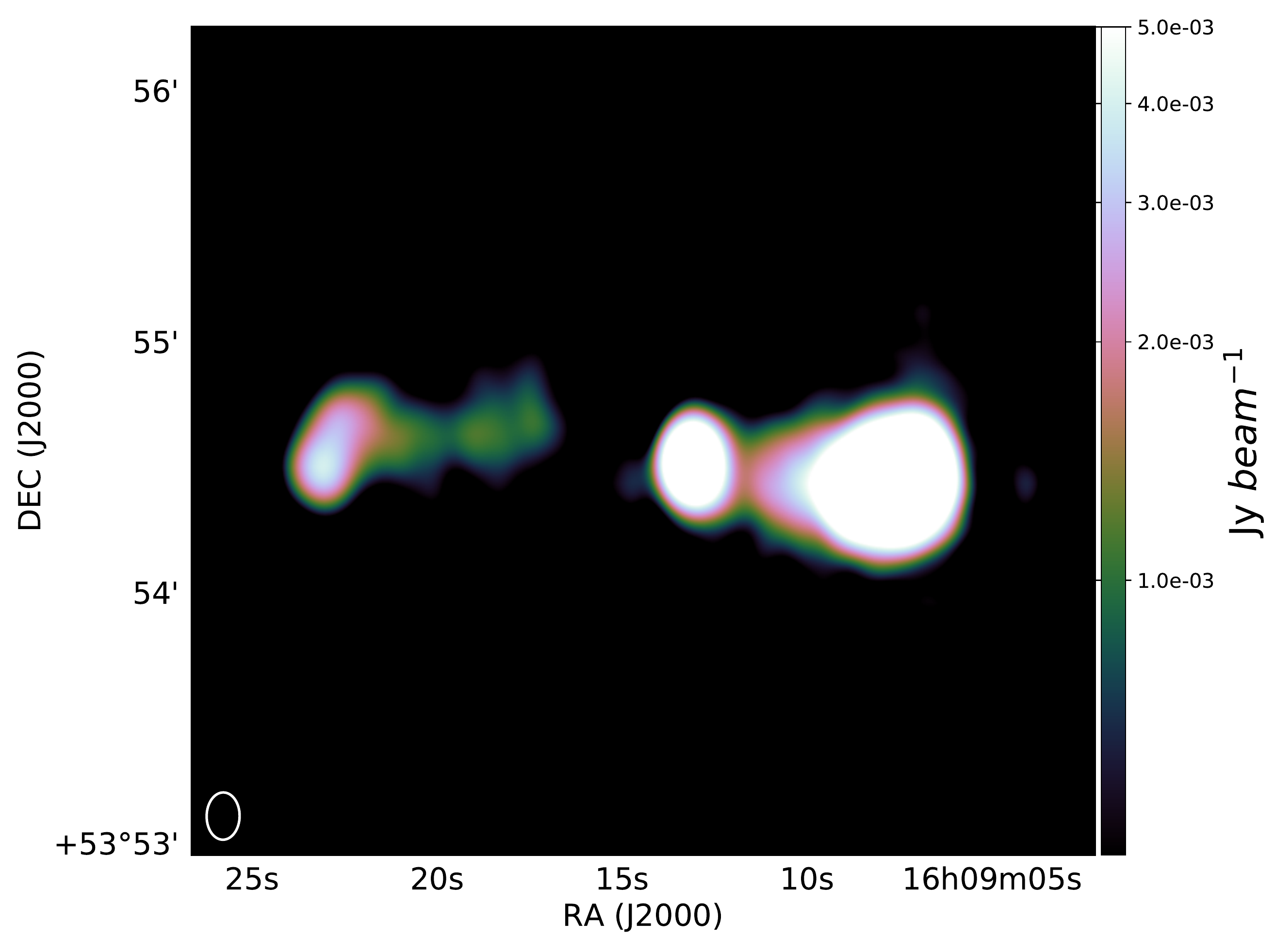}\\
\end{tabular}
\caption{  Left and right panel show some specific sources in the ELAIS-N1 field for direction-independent and -dependent calibration respectively. The distance of sources in each row with respect to the phase center are : $0.22^{\circ}$ (first row), $0.41^{\circ}$ (second row), $0.53^{\circ}$ (third row), $0.62^{\circ}$ (fourth row).  }
\label{fig.source}
\end{figure*}  

    {\bf Calibration and Imaging :} The main pipeline also consists two parts: a direction-independent self calibration part and direction-dependent ionospheric calibration part. Phase-only self-calibration of the target field was started using multi-point source model of the local sky derived from the NVSS catalog. Self-calibration was followed by wide-field imaging and CLEAN deconvolution of the primary beam area and out to 5 primary beam radii to include bright outliers sources to avoid  negative side-lobes of those sources  during imaging. It used Briggs weighting with robust parameter -1, which generally gives well-behaved point spread function (without broad wings) by down weighting the very dense central   $\textit{uv}$-coverage of the GMRT. Phase-only self-calibration were repeated for three more times followed by one round of amplitude and phase self calibration where gain solutions were determined on a longer time-scale than the phase-only solutions. Phase solutions are filtered to separate ionospheric from instrumental effects and instrumental effects were removed from visibilities  (see \citealt{Intema2009}). Between imaging and calibration it constructed residual visibilities by first subtracting  model from data and then Fourier transforming it back to visibility domain.  Then any ripple artifacts in image plane  will show up as a localized and high amplitude peaks in the  $\textit{uv}$-plane and removed those from the data \citep{Intema2016,Intema2014}.

Significant artefacts still remained near bright sources mainly because of residual phase errors due to ionosphere. The gain phases and sky model result from the direction-independent part of the pipeline were sufficient to start direction-dependent (from hereon DD) calibration. DD  gain phases were obtained by peeling bright in-beam sources in the FoV yielding measures of ionospheric phase delay. DD gain phases per time stamp were fitted with a two-layer phase screen model. During imaging of the full FoV, this model was used to  calculate the phase correction per facet  while applying the DD gain tables on fly. At the end of the pipeline, we got the primary beam corrected map of the target field (ELAIS-N1). The final map is shown in Figure \ref{fig.SPAM}. The off-source rms of the map near the phase center is 40 $\mu $Jy/beam  and beam size is 11$\arcsec$ $\times$ 8 $\arcsec$ .

\subsection{ Comparison Between Two Calibration Approaches :} As seen in Figure \ref{fig.SPAM}, there is  significant improvement in dynamic range and there are less  artefacts around bright sources. Although to visualize the improvements after DD calibration, we have shown  four specific sources in ELAIS-N1 field with increasing distance from the phase center in Figure \ref{fig.source}. The left and right columns in  Figure \ref{fig.source} are for direction-independent and -dependent calibration respectively and  from top to bottom source position with respect to phase center is in increasing order. It can be seen that the reconstruction of  the farthest source from phase center (last row of Figure \ref{fig.source})  is very poor in direction-independent calibration in comparison with direction-dependent one. This justifies that for wide FoV and at low-frequency observation direction-dependent calibration is required to reconstruct the sources which get affected due to bright artefacts.  
\begin{figure*}
\centering
\begin{tabular}{lcr}
\includegraphics[width=5.5in,height=4.in]{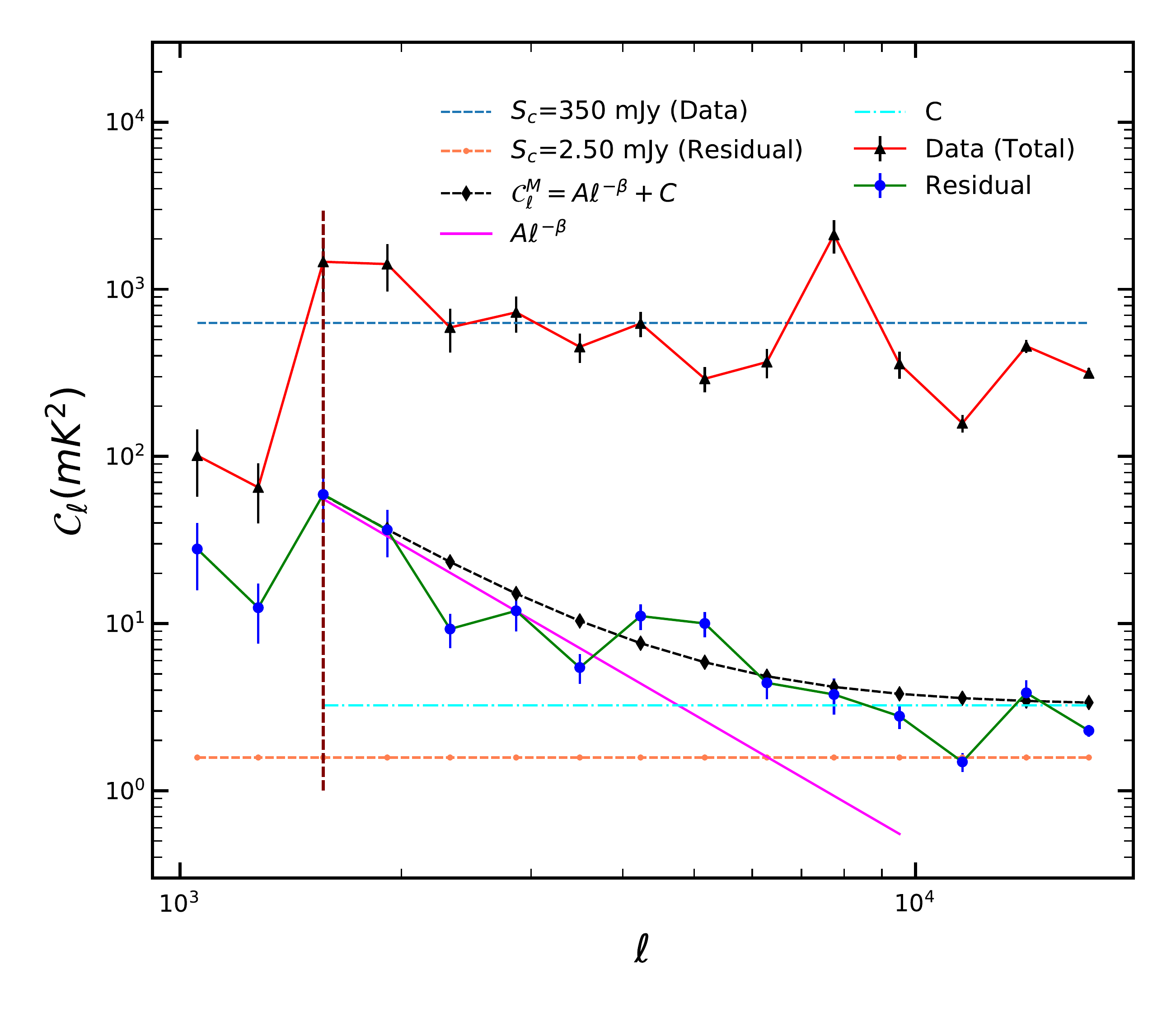} \\
\includegraphics[width=5.5in,height=4.in]{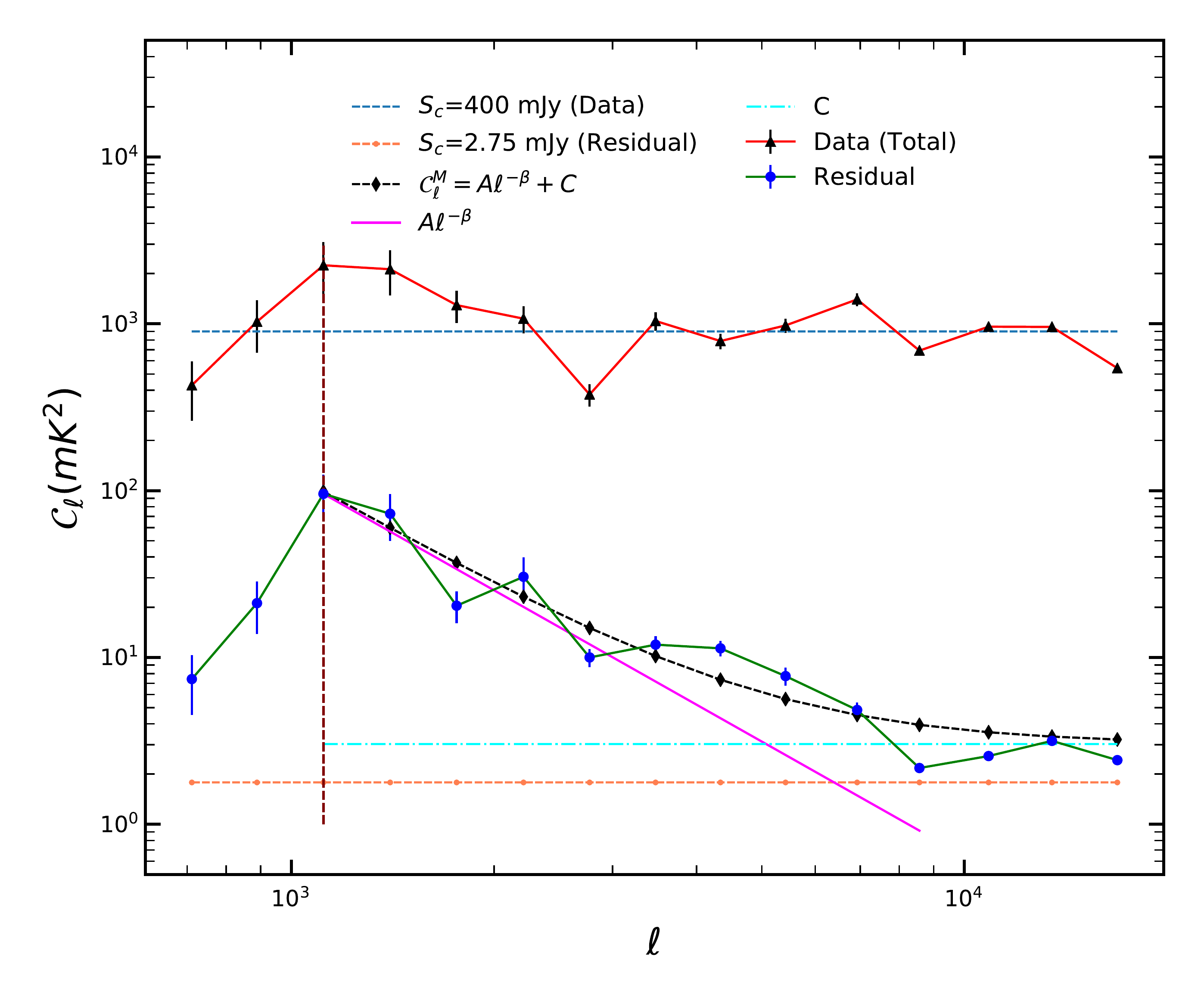}
\end{tabular}
\caption{Estimated angular power spectrum ($C_{\mathcal{\ell}}$) with 1-$\sigma$ error bars before (upper curve in Red) and after source subtraction (lower curve in Green)   with tapering parameter f=0.5 and f=1.0 for direction-independent(top) and -dependent calibrated visibilities (bottom) respectively. The  dashed line (upper horizontal line in Sky) shows foreground contribution due to discrete nature of point sources  based on model prediction of  \citet{Ali2008}. We have excluded the points below the vertical dashed line (in Maroon) from our analysis due to convolution error. The black dashed curve shows the best fitting model, $ C_{\mathcal{\ell}}^{M}= A (1000/\mathcal{\ell})^{\beta} +C $. The dash-dash-dot horizontal line at the bottom shows $C_{\mathcal{\ell}}$ predicted from the residual point sources below a threshold flux density $S_{c} = 2.75mJy$ and $S_{c} = 2.50mJy$ for direction-independent and -dependent calibration respectively. The Magenta line and the horizontal line in cyan represent the power law of the form $A (1000/\mathcal{\ell})^{\beta}$ and the constant term (C) of the best fitted model $ C_{\mathcal{\ell}}^{M}$.}
\label{fig.powerspec}
\end{figure*}

\section{Foreground Characterisation}
\label{TGE_theory}
After making the map with two different calibration approaches we proceed to quantify angular fluctuations in  Galacic and extragalactic foregrounds. To do this we have used Tapered Gridded Estimator(TGE). Here we briefly discuss the basics of TGE and the novelty of this particular estimator, for more details see \citep{Choudhuri2014,Choudhuri2016}. 

\subsection{Tapered Gridded Estimator (TGE) - Brief Background}
TGE uses correlations between gridded visibilities which gives unbiased estimate of angular power spectrum. Spectral smoothness of foregrounds over redshifted HI 21-cm signal holds the promise to extract the cosmological  signal amidst bright foregrounds. But bright sources  at the edge of FoV  can cause of oscillation in the foreground spectra and makes it un-smooth. As a result extraction of the cosmological signal  becomes challenging. Side lobes of these bright sources near the nulls of the primary beam  also causes difficulties to estimate the  power spectrum. TGE overcome these problems by tapering the Primary beam. It cuts off the sky response well before the first null. Another feature of TGE is  it uses gridded visibilities to compute power spectrum. So it  is computationally very fast. This establishes the novelty of TGE to estimate the power spectrum. 

The tapering is incorporated by multiplying the sky with a Gaussian window function $\mathcal{W}(\theta)$=$\exp(\frac{-\theta^{2}}{\theta_{w}^{2}})$ where $\theta_{w}<\theta_{0}$ . As a result sky response falls off well before first null. $\theta_{w}$ is parameterized as $\theta_{w}=f \theta_{0}$, where f is the tapering parameter. Here we have used $\theta_{w}$=44 arcmin with f=1 for direction-dependent calibrated visibilities and $\theta_{w}$=22 arcmin with f=0.5 for direction-independent calibrated visibilities. The reason behind this choice of tapering parameter is discussed in subsection \ref{Robustness}. Both of these are  smaller than FWHM (72$\arcmin$) of the primary beam of  GMRT at 325 MHz.
In the visibility domain  tapering is achieved by  convolving the gridded visibilities in a rectangular gridded plane with the Fourier transform of $\mathcal{W}(\theta)$ - 
\begin{equation}
\mathcal{V}_{cg}=\sum_{i} \tilde{\omega}(\boldsymbol{U_{g}}-\boldsymbol{U_{i}}) \mathcal{V}_{i}
\end{equation}
where $\mathcal{V}_{cg}$ is the convolved visibilities at every grid points g, $\tilde{\omega}$ is the Fourier transform of the tapering window function $\mathcal{W}(\theta)$ and $\boldsymbol{U_{g}}$ refers to the baseline corresponding to the grid points. 
The self-correlation of the gridded and convolved visibilities can be written as - 
\begin{eqnarray}
    \big<|\mathcal{V}_{cg}|^2\big> &=& \Bigg(\frac{\partial B}{\partial T}\Bigg)^2\int d^2U |\tilde{K}(\boldsymbol{U_{g}}-\boldsymbol{U})|^2 \mathcal{C}_{2\pi U_{g}} \\ \nonumber 
    & & + \sum_{i} |\tilde{\omega}(\boldsymbol{U_{g}}-\boldsymbol{U_{i}})|^2 \big<|\mathcal{N}_{i}|^{2}\big>
\end{eqnarray}
where 
\begin{equation}
 \tilde{K}(\boldsymbol{U_{g}}-\boldsymbol{U})= \int d^2U' \tilde{\omega}(\boldsymbol{U_{g}}-\boldsymbol{U'}) B(\boldsymbol{U'}) \tilde{a}(\boldsymbol{U'}-\boldsymbol{U}) 
\end{equation}
is the``gridding kernel" and 
\begin{equation}
B(\boldsymbol{U})=\sum_{i} \delta^{2}_{D}(\boldsymbol{U}-\boldsymbol{U_{i}})
\end{equation}
is the baseline sampling function of the measured visibilities. 
Under the  assumption  $\mathcal{C}_{2\pi U_{g}}$ is nearly constant across the width of $\tilde{K}(\boldsymbol{U_{g}}-\boldsymbol{U})$ we can approximate the convolution as -  
\begin{eqnarray}
    \big<|\mathcal{V}_{cg}|^2\big>&=& \Bigg[\Bigg(\frac{\partial B}{\partial T}\Bigg)^2\int d^2U |\tilde{K}(\boldsymbol{U_{g}}-\boldsymbol{U})|^2 \Bigg]\mathcal{C}_{2\pi U_{g}} + \\ \nonumber 
    & & \sum_{i} |\tilde{\omega}(\boldsymbol{U_{g}}-\boldsymbol{U_{i}})|^2 \big<|\mathcal{N}_{i}|^{2}\big>
\end{eqnarray}
Here again the correlations of tapered gridded visibilities with itself provides an estimate of the angular power spectrum.

The Tapered Gridded Estimator (TGE) is defined as  - 
\begin{equation}
\hat{E_{g}} = M_{g}^{-1} \Bigg( |\mathcal{V}_{cg}|^2 - \sum_{i} |\tilde{\omega}(\boldsymbol{U_{g}}-\boldsymbol{U_{i}})|^2  \mathcal{V}_{i}|^2 \Bigg)
\end{equation}
where $M_{g}$ is the normalizing factor and given as 
\begin{equation}
M_{g}= \Bigg(\frac{\partial B}{\partial T}\Bigg)^2\int d^2U |\tilde{K}(\boldsymbol{U_{g}}-\boldsymbol{U})|^2 - \sum_{i} |\tilde{\omega}(\boldsymbol{U_{g}}-\boldsymbol{U_{i}})|^2  \mathcal{V}_{0} 
\end{equation}

We have calculated $M_{g}$ by using simulated visibilities corresponding to an unit angular power spectrum. The second term in eqn.(6) gives positive noise bias, however, in eqn.(7) this bias is removed by subtracting the auto-correlation of visibilities. 
So , $< \hat{E}_{g}> = \mathcal{C}_{lg}$ gives you the unbiased estimate of the angular power spectrum at the angular multipole $\mathcal{\ell}_{g}= 2 \pi U_{g}$ corresponding to the baseline $U_{g}$.

\subsection{Methodology and Results}
\label{results}
The map of the whole field includes mainly two astrophysical components : extragalctic  point sources and  Galactic diffuse emissions. We first make APS of the total data,i.e, before point source subtraction and then after subtracting those  bright sources  we quantify fluctuations in DGE. We have done this for both calibration approaches and compare the results. Figure \ref{fig.powerspec}  shows the results of the estimated angular power spectrum for direction-independent (top) and direction-dependent (bottom) calibration approaches.

{\bf Point Source contribution :}   The Red curves of both figures show the estimated $\mathcal{C}_{\mathcal{\ell}}$ before point source subtraction. We have found  that for both  calibration processes  the measured $\mathcal{C}_{\mathcal{\ell}}$ is nearly $10^{3} mK^{2}$ across the entire $\mathcal{\ell}$ range considered here.

We have modelled  $\mathcal{C}_{\mathcal{\ell}}$ using the foreground model proposed  in \citet{Ali2008}. The  dashed line in Sky  shows  predicted $\mathcal{C}_{\mathcal{\ell}}$ due to Poisson fluctuations of discrete point sources, where the  flux density of the brightest source in the  direction-independent and -dependent  calibration are $S_{c}= 350 mJy$ and $S_{c}= 40 0 mJy$ respectively. We have found that the estimated $\mathcal{C}_{\mathcal{\ell}}$ before source subtraction across the entire range of angular scales probed here  is nearly flat,  consistent with the model prediction of \citet{Ali2008} .

\begin{figure*}
\centering
\begin{tabular}{cc}
Direction Independent(CASA)  & Direction Dependent (SPAM)\\\\

\includegraphics[width=3.2in,height=2.75in]{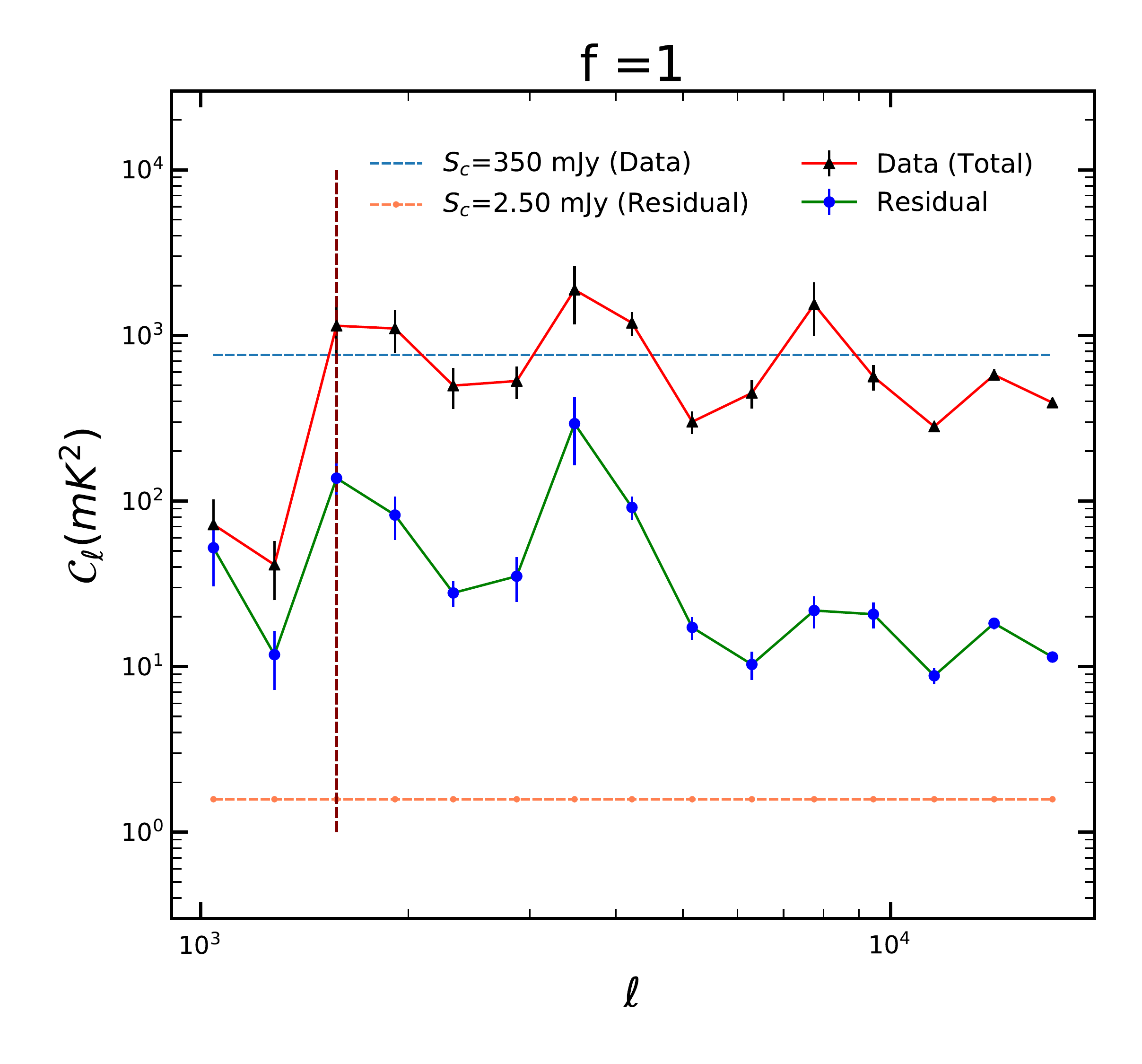}&
\includegraphics[width=3.2in,height=2.75in]{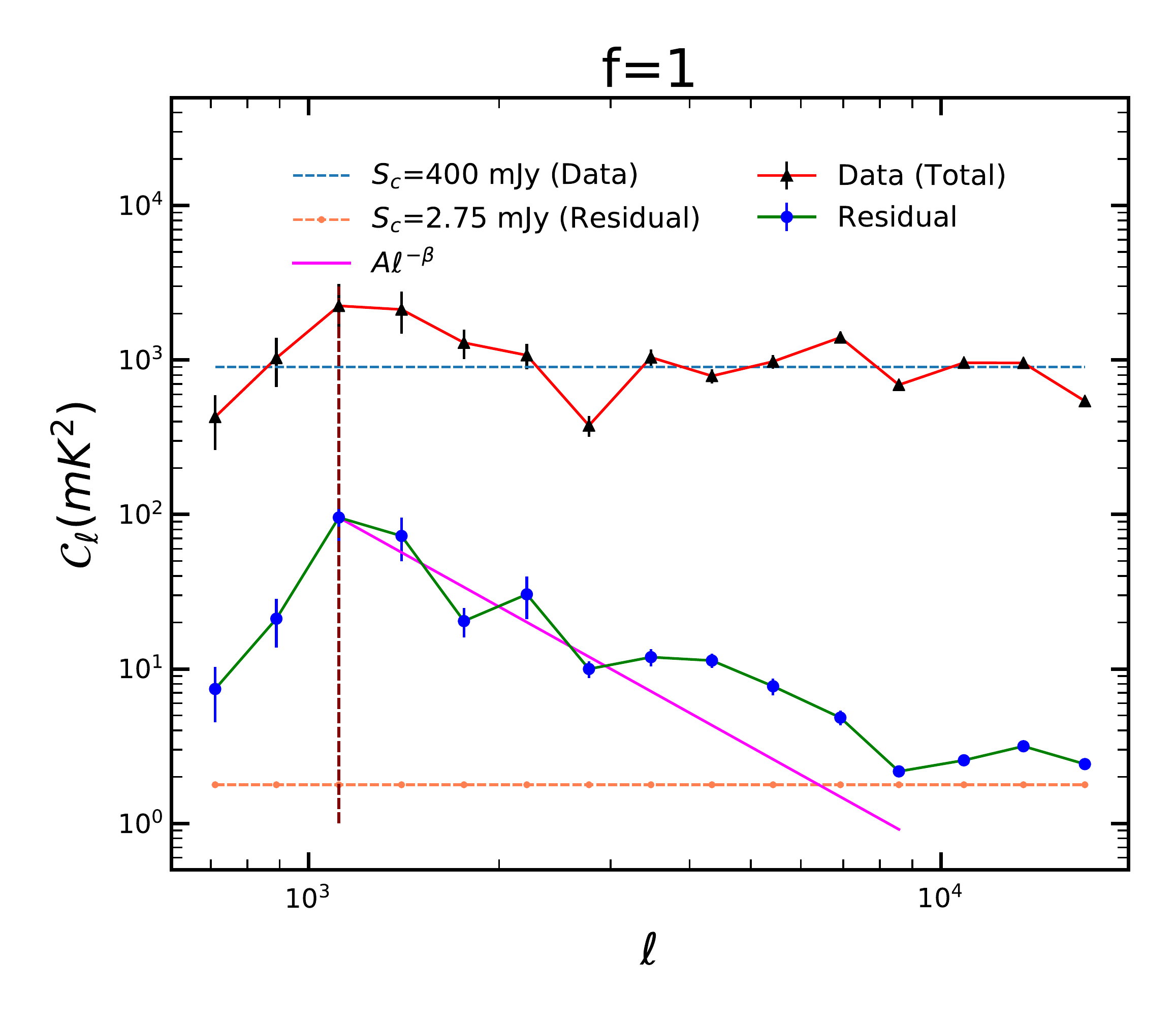}\\
\includegraphics[width=3.2in,height=2.75in]{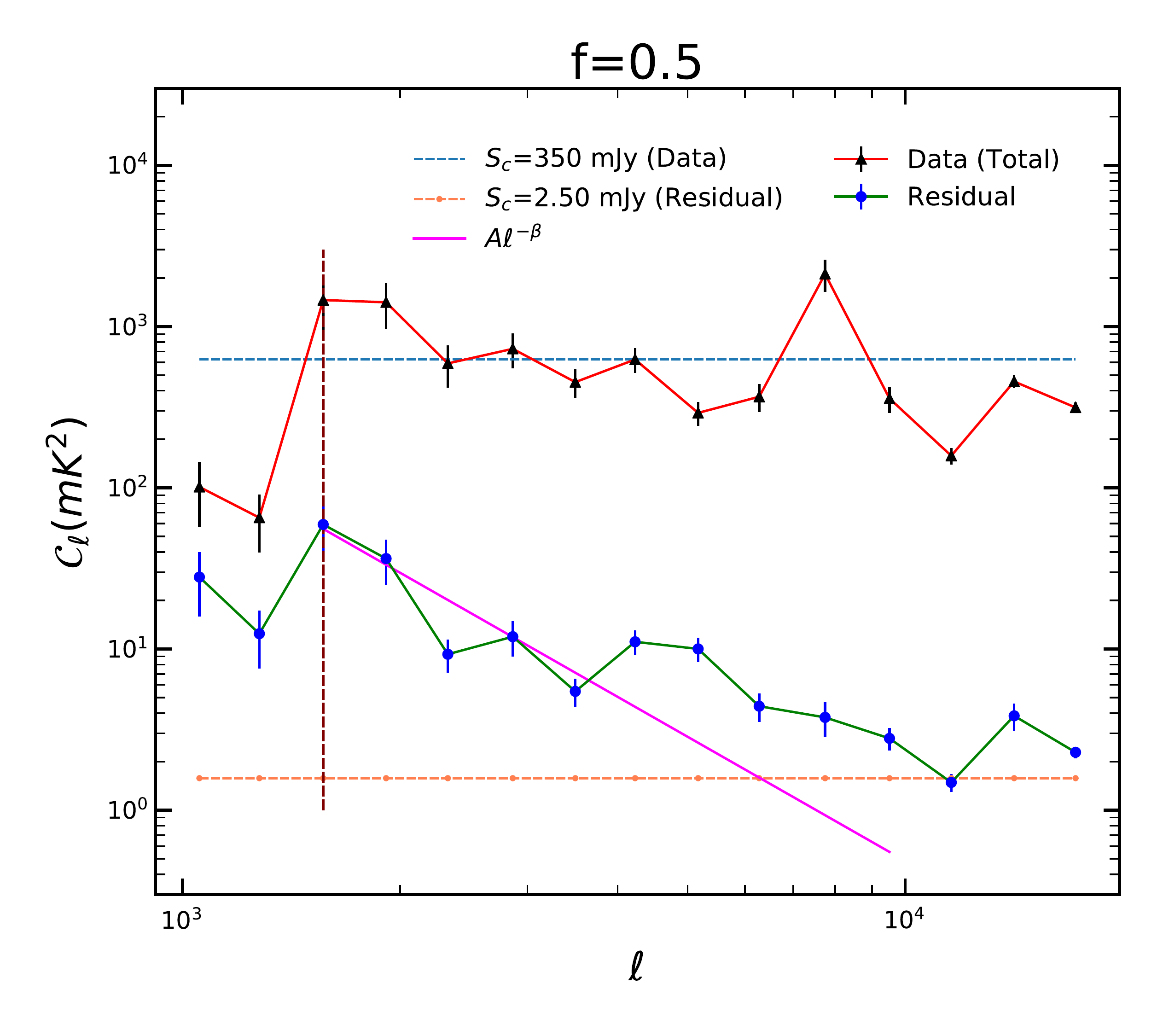}&
\includegraphics[width=3.2in,height=2.75in]{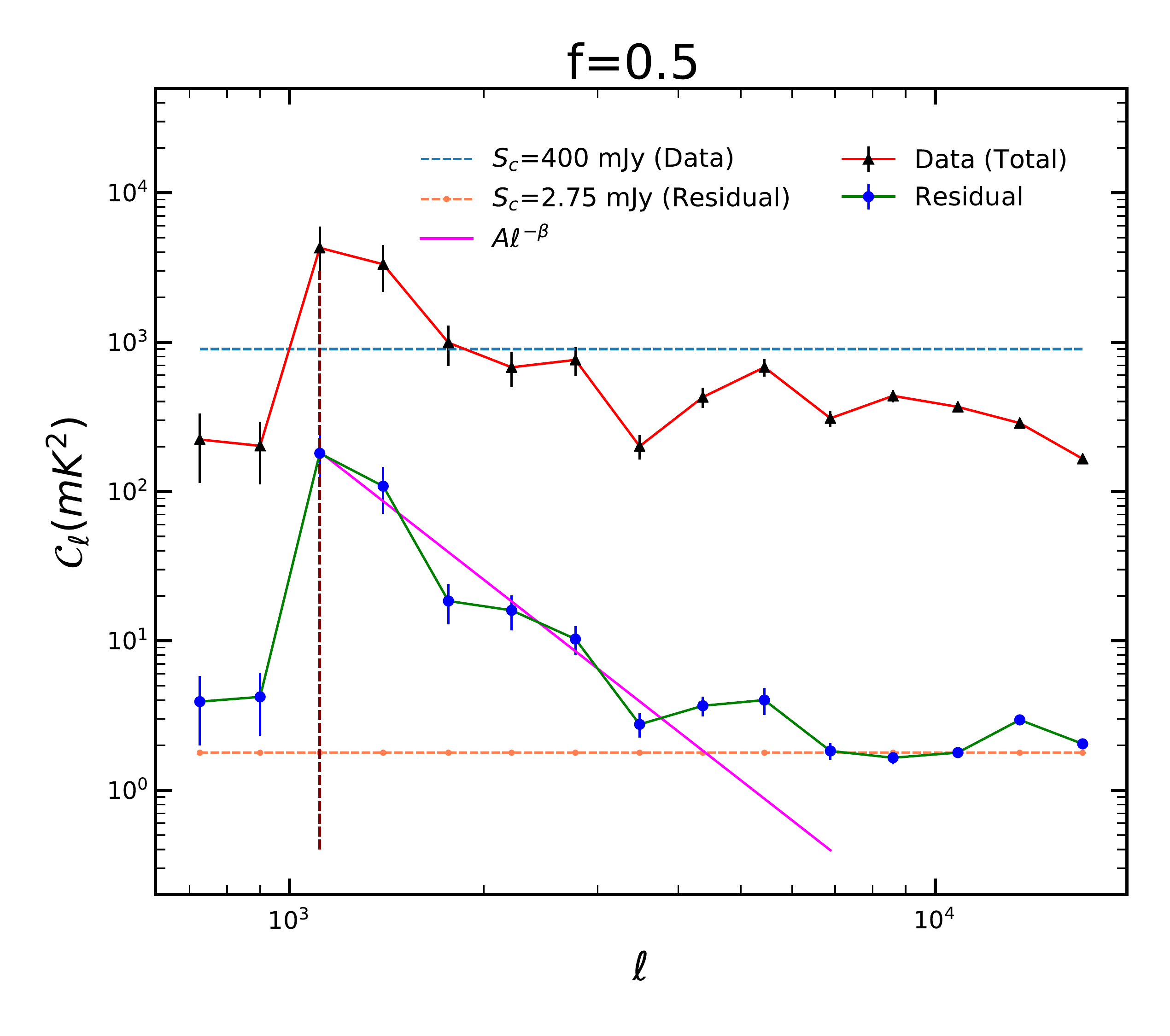}\\
\end{tabular}
\caption{Here we have plotted power spectrum with different tapering parameter(f). The left column shows the estimated power spectrum for direction-independent calibration and right column is for direction-dependent calibration with different tapering parameters f=1.0 (top row) and f=0.5 (bottom row)}
\label{fig.taper}
\end{figure*} 
{\bf DGSE contribution :} We model  the point  sources during CLEANing and subtract that  model from the whole field  using \textit{UVSUB} in CASA. After source subtraction, the residual map   consists of DGSE and residual point sources below the noise level. Angular power spectrum of DGSE is modelled, based on observations,  as a power law of the form $\mathcal{C}_{\mathcal{\ell}}= A \mathcal{\ell} ^{ -\beta}$ and  the Poission fluctuations of residual point sources contributes as a constant term in angular power spectrum \citep{La Porta2008,Ali2008,Ghosh2012,Choudhuri2017}.  In Figure \ref{fig.powerspec}  the Green curve  shows the data points  with $1-\sigma$ error bars of  estimated $\mathcal{C}_{\mathcal{\ell}}$ after point source subtraction. There is significant  drop in power for both the cases after removal of point sources. It is clear from the curve that, $\mathcal{C}_{\mathcal{\ell}}$ shows two different scaling behavior as a function of $\mathcal{\ell}$.  For large angular scales (low $\mathcal{\ell}$), $\mathcal{C}_{\mathcal{\ell}}$ shows  decreasing pattern implying  that  the angular power spectrum is dominated by DGSE. But beyond  certain  $\mathcal{\ell}$ contribution of residual point sources dominates over DGSE and as a result  $\mathcal{C}_{\mathcal{\ell}}$ becomes flat.

The  dash-dash-dot horizontal line in Orange  shows the $\mathcal{C}_{\mathcal{\ell}}$ predicted from the Poisson fluctuations of residual point sources below a threshold flux density of $S_{c} =2.75 mJy$  and $S_{c} =2.50 mJy$ (Figure \ref{fig.powerspec}). Here these high flux densities ($S_{c}$) correspond to bright artefacts remaining in the residual. Except few artefacts the rest of the residual is consistent with noise.  

\begin{table*}
	\centering
	\caption{The table shows the extrapolated values  of the best fitted parameters from different observations at $\mathcal{\ell}=1500$ at our observing frequency for comparative study. }
	\begin{tabular}{l c c c c c c c c c r} 
		\hline
            & Galactic coordinate($\mathcal{\ell}$,b) & $\mathcal{\ell}_{min}$ & $\mathcal{\ell}_{max}$ & A($mK^{2}$) & $\beta$ & $\chi^{2}_{R}$ \\
       \hline
		ELAIS-N1 (Direction-Independent) & ($86.95^{\circ} , +44.48^{\circ}$) & 1565 & 4754 & 62 $\pm$ 6 & 2.55$\pm$0.3  & 1.79 \\
		ELAIS-N1 (Direction-Dependent ) & ($86.95^{\circ} , +44.48^{\circ}$) & 1115 & 5083 &$ 48 \pm 4$ & $2.28\pm 0.4$ & 1.87 \\
        
		\citet{Ghosh2012}  & ($151.8^{\circ} ,+13.89^{\circ}$) & 253 & 800 & $4.16^{a}$ &$2.34 \pm 0.3 $  \\
        
	   \citet{Choudhuri2016} ( DATA 1)  & ($9^{\circ} ,+10^{\circ}$) & 240 & 580 & $2.4^{a}$ &$2.80 \pm 0.3 $ & 0.33\\
       
       \citet{Choudhuri2016} ( DATA 2)  & ($15^{\circ} ,-11^{\circ}$) & 240 & 440 & $0.46 ^{a}$ &$2.2\pm 0.4 $ & 0.15\\
       
        \citet{Bernardi2009} & ($137^{\circ} ,+8^{\circ}$) & 100 & 900 & $2.17^{a}$ &$2.2\pm 0.3$ \\ 
        \citet{Iacobelli2013} & ($137^{\circ} ,+7^{\circ}$) & 100 & 1300 & $-$ &$1.84\pm 0.2$ \\
                               & ($- ,\geq+10^{\circ}$) & - & - & $1.05^{b}$ &$2.88$ \\
                               & ($- ,\leq-10^{\circ}$) & - & - & $1.34^{b}$ &$2.74$ \\
                               & ($- ,\geq+20^{\circ}$) & - & - & $0.5^{b}$ &$2.88$ \\
         \citet{La Porta2008}  & ($- ,\leq-20^{\circ}$) & - & - & $0.3^{b}$ &$2.83$ \\
                               & ($- ,\geq+10^{\circ}$) & - & - & $4.28^{c}$ &$2.80$ \\
                               & ($- ,\leq-10^{\circ}$) & - & - & $4^{c}$ &$2.70$ \\
                               & ($- ,\geq+20^{\circ}$) & - & - & $1.67^{c}$ &$2.83$ \\
                               & ($- ,\leq-20^{\circ}$) & - & - & $0.64^{c}$ &$2.87$ \\
        \hline
	\end{tabular}
\begin{flushleft}
$^a$ Extrapolated from 150 to 325 MHz \\
$^b$ Extrapolated from 1420 to 325 MHz \\
$^c$ Extrapolated from 408 to 325 MHz \\
\end{flushleft}
\label{previous_obs}
\end{table*}
 
{\bf Fitting Routine :} The shortest baseline for direction-independent and -dependent calibration techniques are  U=80$\lambda$  and U=62$\lambda$  corresponding to angular scales of 42$\arcmin$ and 55$\arcmin$ respectively. As a result our observation is not sensitive to intensity variation at angular scales larger than these. Considering the absence of low baselines in the visibility data and taking into account the error introduced by the approximation made during convolution in eqn.(6), we have excluded the $\mathcal{\ell}$ range $\mathcal{\ell} <\mathcal{\ell}_{min}$= 1500 and 1115  for direction-independent and -dependent calibration  respectively.   We have found that for  the whole $\mathcal{\ell}$ range beyond $\mathcal{\ell}_{min}$   a analytical function of the form - 
\begin{equation}
\mathcal{C}_{\mathcal{\ell}}^{M} = A \mathcal{\ell} ^{ -\beta} +C 
\end{equation}

gives the best fit with the reduced $\chi^{2}$ ($\chi^{2}_{R}$) are 1.79 and 1.87 for direction-independent and -dependent calibration approaches respectively.
 The  fitted curves (in Black) are shown in Figure \ref{fig.powerspec}. The best fitted parameters are  (A,$\beta,C$)=($62 \pm 6,2.55 \pm 0.30 ,3.24 \pm 1.09$) and ($ 48 \pm 4, 2.28 \pm 0.4, 3.02 \pm 2.01$) for direction-independent and -dependent calibration respectively. For both the cases we have quoted the value of normalized amplitude at  $\mathcal{\ell} =1500 $.  We have also plotted the power law of the form $\mathcal{C}_{\mathcal{\ell}}= A \mathcal{\ell} ^{ -\beta}$ with best fitted values of (A,$\beta$) (in Magenta)  and the constant term C (in Cyan). It is evident from the plots that residual sources  (the flat part) become dominant over DGSE beyond the intersection point of these two lines at $\mathcal{\ell}$ = 4754 and 5083    for direction-independent and -dependent calibration techniques respectively. So, for DGSE estimated $\mathcal{C}_{\mathcal{\ell}}$ can be well modelled as a power law for $\mathcal{\ell}$ range 1565 $\leqslant \mathcal{\ell} \leqslant 4754 $ and 1115 $\leqslant \mathcal{\ell} \leqslant 5083 $ for direction-independent and -dependent calibration approaches respectively. This steep spectrum is characteristic of fluctuations in DGSE.

 {\bf Other observations :} It is well established from  observations that strength of DGSE is different for different line of sight and for different frequencies. So, it is not justifiable to compare amplitude of angular power spectrum obtained in different observations at different frequencies. Despite this to check consistency,  we have extrapolated the amplitude of $\mathcal{C}_{\mathcal{\ell}}$ obtained from different observations \citep{La Porta2008,Bernardi2009,Ghosh2012,Choudhuri2016} at our observing frequency (325 MHz) at  $\mathcal{\ell} =1500 $ using spectral index $\alpha= -2.5$ ($\mathcal{C}_{\mathcal{\ell}} \propto \nu ^{ 2\alpha}$). The extrapolated values of amplitude together with angular spectral index ($\beta$)  are mentioned in Table \ref{previous_obs}. For all cases, the best fitted  parameter $\beta$ lies within the range of 1.5-3.0 at 150 MHz and higher frequencies.

\subsection{Robustness of the TGE For Direction-Dependent Effects}
\label{Robustness}
We have already mentioned in Sect. \ref{TGE_theory} that novelty of TGE is it tapers the sky response well before first null of the primary beam. The tapering is quantified by the parameter f. Decreasing the value of  f gives higher tapering of primary beam. As a result the effect of bright point sources at the outer region of FoV gets reduced. So at large angular scales  we  expect to get a steep power law pattern in estimated   $\mathcal{C}_{\mathcal{\ell}}$. \cite{Choudhuri2016} has shown with simulated visibility data for GMRT that with increasing tapering of FoV  the fractional deviation of estimated power spectrum from model power spectrum (input of simulation) gets reduced. In other words, this implies that higher tapering gives better result for recovering of input model of angular power spectrum. 

To validate this  we have applied TGE with different tapering parameters (f) and the results are presented in Figure \ref{fig.taper}. We have shown  the estimated $\mathcal{C}_{\mathcal{\ell}}$ with f=0.5 and f=1.0 for both calibration processes and  only the power law fitted line (in Magenta) for clarity. We have found that for direction-dependent calibration, different tapering  gives nearly same results. For both tapering parameters, we have found a range of $\mathcal{\ell}$ where estimated $\mathcal{C}_{\mathcal{\ell}}$ behaves like  a power law and the  fitted parameters are also same within 2$\sigma$ (A=67 $\pm$8, $\beta= 3.0 \pm $0.4, for f=0.5 and A= 48 $\pm$ 4, $\beta=$2.28$\pm $0.39, for f=1.0). 

 In case of  direction-independent calibration,  we have found  some  random fluctuations at certain $\mathcal{\ell}$, for less tapering of sky response with f=1.0 . These fluctuations mainly occur due to bright artefacts at some  localized regions. Increased tapering (f=0.5) however reduced the FoV and  suppressed the effect of those bright sources and we got  a power law pattern  in estimated $\mathcal{C}_{\mathcal{\ell}}$.
Phase only self-calibration failed to suppress the effect of  bright artefacts at large angular distances for direction-independent calibration. So, we need higher tapering of sky response to suppress those effects in estimating angular power spectrum.  Whereas direction-dependent calibration  minimized  the effect of bright sources at large angular scales. So, reducing sky response with different f parameter does not have any significant effect in estimation of  $\mathcal{C}_{\mathcal{\ell}}$.  This effectively validates  robustness  of TGE for unbiased estimation of angular power spectrum from visibility data. 

But due to higher tapering we are unable to recover $\mathcal{C}_{\mathcal{\ell}}$ for low $\mathcal{\ell}$ ( $\leq 1565 $) values in direction-independent calibration approach. Whereas for direction-dependent calibration  we have measured  $\mathcal{C}_{\mathcal{\ell}}$ up to $\mathcal{\ell}$ = 1115. In other words, we have information for large angular scales for direction-dependent calibration technique due to less tapering of FoV. 

\section{Discussion}
\label{conclusion}
We have observed  the ELAIS-N1 field with uGMRT at 325 MHz with main motivation to characterize foregrounds in this field. We have calibrated the visibility data with and without direction-dependent calibration techniques and made two separate  continuum images.  There is significant improvement in dynamic range after direction-dependent ionospheric calibration has been performed and also the imaging artefacts around bright sources have been  minimized.  We have estimated  the angular power spectrum for both direction-independent and -dependent calibrated visibility data sets using TGE. We have found that before source subtraction at most of the $\mathcal{\ell}$ scales probed here estimated $\mathcal{C}_{\mathcal{\ell}}$ is nearly $10^{3}$m$K^{2}$ and remain flat  except at lower $\mathcal{\ell}$ where deconvolution error is significant. We can conclude that the measured $\mathcal{C}_{\mathcal{\ell}}$ is point source dominated and is more than 5-6 orders of magnitude higher than the expected HI signal.

After subtraction of point sources from the entire FoV, the estimated $\mathcal{C}_{\mathcal{\ell}}$ for residual in both cases have shown a steep power law behavior at low $\mathcal{\ell}$ range, which is characteristics of DGSE. We have found a  power law  fit for a specific   $\mathcal{\ell}$ range and the best fitted  amplitude and power law index are (A,$\beta$)= (62,2.55) and (48,2.28) for direction-independent and -dependent calibrations respectively. The slope of angular power spectrum is consistent with the measurements of the previous observations at low frequencies.

We have also estimated the power spectrum using different tapering parameters (f). We have shown that higher tapering of sky response is required to get rid of undesired effects of bright sources in estimating $\mathcal{C}_{\mathcal{\ell}}$   for direction-independent calibration in comparison with direction-dependent one. This validates the robustness of TGE  with real data. This fact is well established in \cite{Choudhuri2016} for simulated visibility data for GMRT.

Analytic estimates of the HI signal shows that at low frequencies the amplitude of HI signal is nearly $10^{-1} mK^{2}$, which is very feeble in comparison with bright foregrounds. After point source subtraction the estimated $\mathcal{C}_{\mathcal{\ell}}$ for DGSE is still 2-3 orders of magnitude higher than the expected HI signal. Proper modelling of point sources and perfect subtraction from the data is very crucial to extract this faint HI signal.  But as foreground spectrum is smooth in comparison to HI signal, it does not decorrelate faster than the HI signal with frequency. This fact is the key to extract the faint HI signal from strong foregrounds. 

The measured $\mathcal{C}_{\mathcal{\ell}}$ for DGSE is different for different patch of sky and at different frequencies. We have analyzed the ELAIS-N1 field to characterize the foregrounds at sub-degree angular scales. The characterization of foregrounds will help us to extract HI signal when the data will be available from upcoming experiments like HERA, PAPER, SKA-low, SKA-mid etc. But it is evident from Table \ref{previous_obs} that only few observations are available across different patches of sky at these low frequencies. For those, the corresponding angular power spectrum for foregrounds are available in literature. In order to constrain the foreground across wide patches of the sky, there is a need to extend sensitive low frequency observations at different patches of the sky. In our second paper, we will present the analysis for a wider bandwidth data (200 MHz). However, even with this limited bandwidth, we have been able to constrain the foreground angular power spectrum and demonstrate the efficiency of TGE against direction-dependent effects.

We also plan to extend this work for other well-known target fields at low frequency with continuum observation to find out the variation in the nature of DGSE.
\newline

{\bf ACKNOWLEDGEMENTS}

We  thank the staff of GMRT for making this observation possible. GMRT is run by National Centre for Radio Astrophysics of the Tata Institute of Fundamental Research. AC would like to thank DST for INSPIRE fellowship. AC thanks Ramij Raja, Anirban Roy, Bhargav Vaidya and Manoneeta Chakraborty for helpful discussion. NR acknowledges support from the Infosys Foundation through the Infosys Young Investigator grant. We warmly thank the  anonymous  referee  for  helpful  comments.







\bsp	
\label{lastpage}
\end{document}